\PassOptionsToPackage{dvipsnames,svgnames,x11names}{xcolor}
\documentclass[a4paper,aps,preprint,nofootinbib,floatfix,superscriptaddress,prd]{revtex4-2}
\usepackage{graphicx,xcolor}
\usepackage{booktabs}
\usepackage{amsmath,amssymb,mathtools}
\usepackage{dsfont}
\usepackage[pdfusetitle,pdfencoding=auto]{hyperref}
\DeclareMathOperator{\Tr}{Tr}
\hypersetup{
    colorlinks,
    linkcolor={red!50!black},
    citecolor={blue!50!black},
    urlcolor={blue!80!black}
}

\newcommand{\mpl}{M_{\rm Pl}}

\newcommand{\calL}{{\cal L}}
\newcommand{\calO}{{\cal O}}

\begin{document}

\title{A two scalar triplets model as common origin for dark matter, neutrino masses, baryon asymmetry and inflation}

\author{Sin Kyu Kang}
\email{skkang@seoultech.ac.kr}

\affiliation{%
School of Natural Science, Seoul National University of Science and Technology,
232 Gongneung-ro, Nowon-gu, Seoul, 01811, Korea
}
\author{Raymundo Ramos}
\email{raramos@kias.re.kr}
\affiliation{%
Quantum Universe Center, Korea Institute for Advanced Study, Seoul 02455, Korea
}

\date{\today}

\begin{abstract}
\noindent 
We propose an extension of the standard model (SM)
by two $SU(2)$ triplet scalars and an inert $SU(2)$ doublet.
We demonstrate that this setup
can simultaneously
produce an inflaton
and baryon asymmetry in the early universe,
provide a dark matter candidate
and explain the smallness of neutrino masses.
The two triplets are particularly important
as they become mediators for the production of dark matter
and the generation of lepton asymmetry,
as well as contribute an inflaton.
The inert doublet results in a dark matter candidate.
The required $CP$-violation for lepton asymmetry is obtained
by interference between the triplet mediators that communicate
the dark sector to the SM sector.
More precisely, the complex Breit-Wigner propagators of the triplets
and their mixing, result in an asymmetric production of leptons and antileptons
that is boosted before dark matter freeze-out.
In this case, simultaneously achieving enough dark matter relic abundance
and proper matter-antimatter asymmetry limits the available parameter space of the model.
Moreover, the scalar triplets are coupled non-minimally to gravity and give rise to the inflaton.
We calculate the inflationary parameters and check that we can obtain predictions consistent with Planck constraints from 2018.
We also perform an analysis of the reheating for the inflaton decays/annihilations to relativistic SM particles.

\end{abstract}

\maketitle
\newpage

\section{Introduction}
\label{sec:intro}

The quest to describe the properties of dark matter (DM) has required considerable efforts both on the theoretical and experimental sides.
Even though DM has for the moment escaped all experiments that attempt to describe its particle properties,
observations of the cosmic microwave background radiation have allowed an estimation of its abundance.
From Planck satellite data published in 2018,
we know that the universe is composed by 5 times more DM than baryonic matter~\cite{Planck:2018vyg}.
The main paradigm to describe DM in particle physics is the well studied weakly interacting massive particles (WIMPs).
This paradigm is characterized by a DM candidate that,
besides gravitational interaction,
has interaction rates of similar size or smaller than those associated with the weak force.
This type of DM is produced thermally by selfannihilation via the aforementioned interactions.
The existence of DM is unexplained in the standard model (SM) of particle physics,
as well as the presence of baryon asymmetry in the universe~\cite{Sakharov:1967dj}.
A well established mechanism to explain baryon asymmetry is the out-of-equilibrium decays of heavy particles~\cite{Weinberg:1979bt,Kolb:1979qa},
for example, through leptogenesis~\cite{Fukugita:1986hr},
by generating leptonic asymmetry that is then converted into baryon asymmetry by sphaleron transitions.
One advantage of this scenario is the employment of the seesaw mechanism,
which in turn also explains the origin of tiny neutrino masses~\cite{ParticleDataGroup:2022pth}.
In a recent proposal,
a different mechanism for non-zero asymmetry
is formulated via interference of $2 \to 2$ scattering
with unstable mediators~\cite{Dasgupta:2019lha}.
In this case, complex couplings
and the Breit-Wigner complex propagators,
involved in the construction of the scatterings,
are responsible for $CP$-asymmetry.
Moreover, if one side of the scatterings involves a DM candidate pair,
we can achieve an interesting correlation between the generation of baryon asymmetry via leptogenesis
and the production of DM relics.

In the present work, we expand on the idea presented in Ref.~\cite{Dasgupta:2019lha}
by considering the mixing of propagators in $2 \to 2$ scatterings,
which provides an additional source of $CP$-asymmetry beyond the tree-level interference mechanism discussed in Ref.~\cite{Dasgupta:2019lha}.
We introduce two $SU(2)$ scalar triplets
with Yukawa couplings to the leptonic sector,
where the couplings carry a relative non-vanishing complex phase.
Tiny neutrino masses are generated via the inverse type-II seesaw mechanism~\cite{Li:1985hy,Lusignoli:1990yk,deSPires:2005yok,Freitas:2014fda,deSousaPires:2018fnl}.
For the DM component,
we add an inert $SU(2)$ doublet stabilized by a $Z_2$ symmetry,
leading to a phenomenology analogous to the inert Higgs doublet model~\cite{Deshpande:1977rw,LopezHonorez:2006gr,Gustafsson:2010zz,Arhrib:2013ela,Belyaev:2016lok,Fan:2022dck}.
By allowing the triplets to interact simultaneously with the DM candidate and leptons,
we obtain annihilation scatterings mediated by the triplet components,
whose propagators mix at the one-loop level, producing $CP$-asymmetric annihilations.
This one-loop propagator mixing has been adapted from a previous proposal employing triplets~\cite{Hambye:2000ui}
and implemented in conjunction with the approach of Ref.~\cite{Dasgupta:2019lha}.
While Ref.~\cite{Kang:2023iur} already explored the same scalar content, our work goes beyond the earlier scheme by showing that loop-induced propagator mixing provides a distinct source of CP violation beyond the tree-level interference, which plays an essential role in leptogenesis.
At the same time, the model consistently incorporates inflation and reheating, embedding the early universe dynamics within a unified setup. Most importantly, in contrast with conventional type-II leptogenesis based on heavy-triplet decays, our scenario establishes a direct dynamical connection between the generation of lepton asymmetry and the freeze-out of DM through scattering processes. This feature renders baryogenesis predictive within the same parameter space that controls the relic abundance, and we demonstrate its viability by presenting benchmark points that reproduce both the observed relic density and the baryon asymmetry

In our model, the scalar triplets not only mediate DM interactions and generate lepton asymmetry, but also act as inflaton candidates through non-minimal couplings to gravity,
thus forming part of interconnected aspects of a unified scalar framework.
We calculate the corresponding inflationary parameters
and verify consistency with the latest Planck 2018 constraints,
and we further analyze the reheating stage after inflation,
focusing on inflaton decays and scatterings into relativistic SM particles.

The rest of this paper is organized as follows:
in Sec.~\ref{sec:tripidm} we describe the main details of extending the SM with an inert doublet and two triplets,
in Sec.~\ref{sec:tripneu} we present the contributions from the Yukawa couplings between triplets and leptonic doublets,
in Sec.~\ref{sec:asymmetry} the origin of asymmetry in this model and its relation to DM is demonstrated,
finally, in Sec.~\ref{sec:conclusion} we summarize and conclude.
Other relevant details about the scalar potential of the model are given in Appendix~\ref{app:scalars}.

\section{Extension by triplets and inert doublet}
\label{sec:tripidm}

The extension of the SM used in this work is the same as the one presented in Ref.~\cite{Kang:2023iur}.
In that work, the SM is extended with two scalar triplets of $SU(2)$ and an inert $SU(2)$ doublet.
Additionally, a $Z_2$ symmetry is assumed under which the SM sector and the triplets
have even charges while the inert doublet has odd charge.
This detail eventually will stabilize the DM candidate that results from the inert doublet.
The scalars have the following decomposition
\begin{equation}
    \label{eq:scalars}
	\Phi_1 = \begin{pmatrix}
		0 \\
		\frac{1}{\sqrt{2}}(h_1 + v)
	\end{pmatrix},\quad
	\Phi_2 = \begin{pmatrix}
		\Phi_2^+ \\
		\frac{1}{\sqrt{2}}(H^0 + i A^0)
	\end{pmatrix},\quad
	\Delta_{n} = \begin{pmatrix}
		\delta^+_{n}/\sqrt{2} & \delta^{++}_{n} \\
        \delta^0_{n} + u_n/\sqrt{2} & -\delta^+_{n}/\sqrt{2}
	\end{pmatrix},
\end{equation}
where $\Phi_1$ is the SM-like $SU(2)$ doublet,
$\Phi_2$ is the inert $SU(2)$ doublet,
and $\Delta_n$ with $n\in \{1,2\}$ are two additional $SU(2)$ triplets.
The terms $v$ and $u_n$ represent the vacuum expectation values (VEV) of the scalars.
We will follow a notation where the VEVs of the triplets
are represented by
$u_1 = u \cos\beta$ and $u_2 = u \sin\beta$, with $u = \sqrt{u_1^2 + u_2^2}$~\cite{Ferreira:2021bdj}.
The part of the scalar potential that is relevant for the discussion of this work is
\begin{equation}
    \label{eq:tripletpot}
    V \supset \sum_{n=1}^2 \left[ M_n^2 \Tr\left(\Delta_n^\dagger \Delta_n\right) +
    \left(\sum_{m=1}^2 \mu_{nm} \Phi_m^T i\sigma^2\Delta_n^\dagger \Phi_m + \text{H.c.}\right)\right],
\end{equation}
where we can interpret the second part, the terms between parenthesis,
as terms that softly break a global $U(1)$ symmetry in the scalar potential.
Minimization of the potential yields the following relationships
\begin{equation}
    M_1^2 \approx \frac{\mu_{11} v^2}{\sqrt{2} u \cos\beta}\,,\quad
    M_2^2 \approx \frac{\mu_{21} v^2}{\sqrt{2} u \sin\beta}.
\label{Masses}
\end{equation}
As mentioned earlier, the terms with $\mu_{nm}$ in the potential of Eq.~\eqref{eq:tripletpot}
softly break a $U(1)$ global symmetry and can be assumed to be small.
Assuming that the triplet states have masses of $\mathcal{O}(1\ \text{TeV})$ or higher
and that the $\mu_{n1}$ parameters are small for the reasons above,
we obtain that $u$ has to be much smaller than $v$.
In the discussion below we will see that this is convenient for generating small neutrino masses.
Additionally, the $\mu_{n2}$ parameters affect the communication between
dark sector and SM sector via triplet mediators and their smallness
guarantees that this connection is kept weak.
A full description of the complete scalar potential,
its minimization and
details about diagonalization of scalars
is given in the Appendix of Ref.~\cite{Kang:2023iur}
and summarized in Appendix~\ref{app:scalars} of this work.

\subsection{Leptonic sector and neutrino masses}
\label{sec:tripneu}

\subsubsection{Yukawa couplings}
\label{sec:tripneuyuk}

As mentioned earlier,
the triplets are responsible for generating tiny neutrino masses.
The mechanism that allows this is known as inverse type-II seesaw~\cite{Li:1985hy, Lusignoli:1990yk,deSPires:2005yok,Freitas:2014fda,deSousaPires:2018fnl}.
The Yukawa terms coupling the scalar triplets and the left-handed $SU(2)$ doublets are
\begin{equation}
\label{eq:LagYuk}
    - \mathcal{L}_\mathrm{Yuk} =
        \sum_{n=1}^2 \sum_{j,k} Y^{\Delta_n}_{jk} L_j^T \mathcal{C}^\dagger i \tau_2 \Delta_n L_k + \mathrm{H.c.}\,,
\end{equation}
where the flavors are indicated by indices $j,k$, $C$ represents the charge conjugation matrix,
and $\tau_2$ is the second Pauli matrix.
The Yukawa couplings, $Y^{\Delta_n}_{jk}$,
are taken as complex and
can be considered part of $3\times 3$ matrices $Y^{\Delta_n}$.
To avoid dangerous flavor changing neutral currents~\cite{Pich:2009sp}
we apply an alignment condition $Y^{\Delta_2} = \xi Y^{\Delta_1}$,
where $\xi$ is taken complex.
Defining $Y^\Delta \equiv Y^{\Delta_1} = Y^{\Delta_2} \xi^{-1}$ we obtain
\begin{equation}
\label{eq:LagYuk2}
  - \mathcal{L}_\mathrm{Yuk} =
    \sum_{j,k} Y^{\Delta}_{jk} L_j^T \mathcal{C}^\dagger i \tau_2 \left(
      \Delta_1 + \xi \Delta_2
    \right) L_k + \mathrm{H.c.}
\end{equation}

\subsubsection{Neutrino masses and mixing}
\label{sec:neutmass}

When the triplets fall to the minimum of the potential and acquire VEVs,
the Yukawa couplings result in masses for the neutrinos
that can be arranged in a $3\times 3$ matrix as follows
\begin{equation}
\label{eq:mnuyukawa}
  M^{\nu}_{jk} =
      \sqrt{2} Y^{\Delta}_{jk} u \cos\beta \left( 1 + \xi \tan\beta \right)\,,
\end{equation}
where we can readily see that, without any assumptions about other parameters,
having $u \ll v$ as mentioned in Sec.~\ref{sec:tripidm},
it can effectively set the scale for neutrino masses.
When we assume that the charged lepton masses are diagonal,
this matrix can be diagonalized by the Pontecorvo-Maki-Nakagawa-Sakata (PMNS) matrix, $U_\text{PMNS}$, as
\begin{equation}
\label{eq:upmnsdiag}
	U_\text{PMNS}^T M^\nu U_\text{PMNS} = \text{diag}(m^\nu_1, m^\nu_2, m^\nu_3) \equiv M^\nu_d,
\end{equation}
where the $U_\text{PMNS}$ matrix can be parameterised as~\cite{ParticleDataGroup:2024cfk}
\begin{equation}
\label{eq:upmns}
U_\text{PMNS} = \left(
\begin{array}{ccc}
    c_{12} c_{13} & s_{12} c_{13} & s_{13} e^{-i\delta} \\
    -s_{12} c_{23} - c_{12} s_{13} s_{23} e^{i\delta} &  c_{12} c_{23} - s_{12} s_{13} s_{23} e^{i\delta} & c_{13} s_{23} \\
     s_{12} s_{23} - c_{12} s_{13} c_{23} e^{i\delta} & -c_{12} s_{23} - s_{12} s_{13} c_{23} e^{i\delta} & c_{13} c_{23}
\end{array}
\right)\times P
\end{equation}
with $P=\mathrm{diag} (1, e^{i\phi_1/2}, e^{i\phi_2/2})$.
To simplify the numerical work, we use Eqs.~\eqref{eq:mnuyukawa} and~\eqref{eq:upmnsdiag}
to obtain
\begin{equation}
\label{eq:yukawa_upmns}
  \sqrt{2} Y^{\Delta}_{jk} u \cos\beta \left( 1 + \xi \tan\beta\right) =
      \left(U^*_\text{PMNS} M^\nu_d U^\dagger_\text{PMNS}\right)_{jk}\,.
\end{equation}
where the left-hand side contains all the theoretical parameters.
After fixing the lightest neutrino mass and the two phases, $\phi_1$ and $\phi_2$,
the rest of the parameters in the right-hand side
can be fixed from neutrino oscillation measurements.

\begin{figure}[tb]
    \center
    \includegraphics[width=0.5\textwidth]{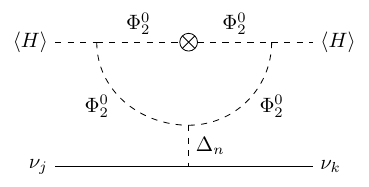}
    \caption{\label{fig:massloop}%
        Feynman diagram for the 1-loop contribution to neutrino masses.
    }
\end{figure}

Additionally,
couplings between $h_1$ and two $H^0$ via the $\lambda_5$ parameter,
make possible the generation of 1-loop corrections to neutrino masses
via the diagram of Fig.~\ref{fig:massloop}.
We make an estimation of the correction from this diagram to neutrino masses
by reducing the loop to an effective three-vertex
between two Higgs doublets and one triplet.
Since this is analogous to the coupling $\mu_{n1}$ in the scalar potential,
we name this effective coupling $\mu'_{n1}$, given by the following expression
\begin{equation}
    \mu_{n1}' = \frac{1}{32 \pi^2}\mu_{n2}\lambda_A^2 \lambda_5 v^4 f(m^2_{\Phi_2^0})
\end{equation}
where the effect of the loop correction is contained in the function $f$ that involves
integration of the propagators in the loop times a factor for the mass insertion.
For the mass of the particle running in the loop we use the placeholder $m^2_{\Phi_2^0}$.
We can use the benchmark values of Table~\ref{tab:benchmarkpts}
to make an estimation of the value of this effective vertex.
We obtain for $v^4/m^4_{\Phi_2^0}\sim 2\times10^{-4}$
and for $\mu_{n2}\lambda_A^2\lambda_5 \sim 7\times 10^{-9}$~GeV at the largest.
Note that in Table~\ref{tab:benchmarkpts} the masses of the pseudoscalar and inert Higgs
are nearly degenerated.
As this estimation tends to overestimate the value of the effective coupling,
and since the effective coupling is much smaller than the tree level $\mu_{n1}$,
we conclude that the value of the loop correction is negligible
and therefore ignore it.

\subsubsection{Couplings between leptons and scalars}
\label{sec:lepscouplings}

The couplings between leptons and components of the triplets
are of particular importance to leptogenesis and DM evolution.
From Eq.~\eqref{eq:LagYuk}, such couplings are described
by the following terms in the potential
\begin{equation}
    \mathcal{L}_\mathrm{Yuk} \supset
        Y^\Delta_{jk} \left[
            - \bar{\nu}^C_j \nu_k \left(\delta^0_1 + \xi \delta^0_2\right)
            + \sqrt{2} \bar{\nu}^C_j \ell_k \left(\delta^{+}_1 + \xi \delta^{+}_2\right)
            + \bar{\ell}^C_j \ell_k \left(\delta^{++}_1 + \xi \delta^{++}_2\right)
        \right]
        + \text{H.c.}
\end{equation}
Considering these couplings
together with the coupling between $\Phi_2$ and the triplets in Eq.~\eqref{eq:scalars},
we obtain $2 \to 2$ scatterings between the dark sector and the leptonic sector.
Note that, in this case, the components of the triplets, $\Delta_1$ and $\Delta_2$,
are the mediators for these $2 \to 2$ scatterings.
Their relevance for leptogenesis and DM production will be detailed in the next section.

\section{Origin of asymmetry}
\label{sec:asymmetry}

\begin{figure}[tb]
    \center
    \includegraphics[width=0.5\textwidth]{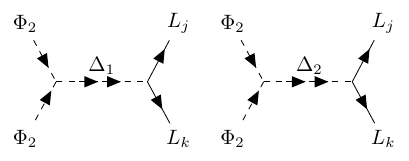}
    \includegraphics[width=0.5\textwidth]{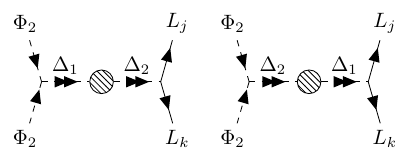}
    \caption{\label{fig:schan}%
        Feynman diagrams for processes mediated by $\Delta_n$
        with interference that contributes to matter asymmetry.
    }
\end{figure}

The generation of lepton asymmetry is, initially,
based on the approach of Ref.~\cite{Dasgupta:2019lha},
where CP violation originates from the tree-level interference between $s$-channel scattering processes
with unstable mediators.
In the present work, we go beyond this setup by showing that the
$2 \to 2$ processes responsible for the asymmetry
receive additional and genuinely new contributions from the
one-loop mixing of scalar triplet propagators.
These processes arise from the trilinear couplings of Eq.~\eqref{eq:tripletpot}
between the triplets and $\Phi_2$, together with the Yukawa couplings of triplets
to leptons in Eq.~\eqref{eq:LagYuk2}.
The corresponding Feynman diagrams are shown in Fig.~\ref{fig:schan}.
In contrast to Ref.~\cite{Dasgupta:2019lha},
where such loop-induced propagator effects were absent,
and also to Ref.~\cite{Kang:2023iur},
which considered only tree-level diagrams in the same scalar setup,
we explicitly incorporate the one-loop propagator mixing between $\Delta_1$ and $\Delta_2$.
This represents a qualitatively new source of CP violation,
since the interference involves not only different mediators but also their loop-induced mixing structure.

The two upper diagrams correspond to tree-level
contributions and were also considered in Ref.~\cite{Kang:2023iur}.
The two lower diagrams are the result of
mixing between the components of $\Delta_1$ and $\Delta_2$
due to their couplings to the same pairs of particles.
Following the formalism of Ref.~\cite{Hambye:2000ui},
we can describe this mixing using the effective terms
\begin{equation}
\Delta^{\dagger}_a \left(M_D^2\right)_{ab} \Delta_b+h.c.
\end{equation}
with the mass matrix $M_D$ given by
\begin{equation}
\label{eq:denmat}
M_D^2=\left(
\begin{array}{cc}
M^2_{\Delta_1}-i C_{11} & -i C_{12} \\
-i C_{21} & M^2_{\Delta_2}-i C_{22}
\end{array}
\right)
\end{equation}
where
\begin{equation}
    \label{eq:cab}
C_{ab}=\Gamma_{ab}M_{\Delta_b} =\frac{1}{8\pi}\left( \mu_{a2}\mu^{\ast}_{b2} + p^2\sum_{\alpha\beta}Y^{\Delta_a \ast}_{\alpha\beta}
Y_{\alpha\beta}^{\Delta_b} \right)\,.
\end{equation}
We have ignored contributions from the SM Higgs scalars because they are much smaller than those we keep.
For the moment we will stay in this basis to contrast changes due to one-loop corrections.

From Ref.~\cite{Dasgupta:2019lha} we know that to generate a non-zero asymmetry factor
we need complex Yukawa couplings, a relative non-zero phase from $\xi$
and non-zero decay widths for the propagators
to achieve asymmetry from interferences.
These elements and the one-loop corrections from the discussion above
result in the $CP$-asymmetry factor
\begin{align}
    \delta \equiv {}& |\mathcal{M}|^2 - |\bar{\mathcal{M}}|^2 \nonumber \\
    = {}&  -4 \left[
         \left|Y^\Delta_{jk}\right|^2\mathrm{Im}\left[\mu_{12}\mu_{22}^*\xi^*\right]
          \mathrm{Im}\left[\frac{1}{\mathcal{S}_1 \mathcal{S}_2^{\ast} }\right]\right. \nonumber \\
    &+  
        \frac{\left|Y^\Delta_{jk}\right|^2}{\left|\mathcal{S}_1 \right|^2}
        \left(\left|Y^\Delta_{lm}\right|^2+\left|\mu_{12}\right|^2\right)\mathrm{Im}\left[\mu_{12}\mu_{22}^*\xi^*\right]
          \mathrm{Im}\left[\frac{C_{12}^{\ast}}{\mathcal{S}_2^{\ast}} \right] \nonumber \\
    &+  \left.
         \frac{\left|Y^\Delta_{jk}\right|^2}{\left|\mathcal{S}_2 \right|^2}
       \left(\left|\xi Y^\Delta_{lm}\right|^2+\left|\mu_{22}\right|^2\right)\mathrm{Im}\left[\mu_{12}^* \mu_{22} \xi \right]
          \mathrm{Im}\left[\frac{C_{21}^{\ast}}{ \mathcal{S}_1^{\ast}} \right] \right]
|\mathcal{W}|^2\,, \label{eq:cpasymfac}
\end{align}
where we have used $\mathcal{S}_i^{-1}=s-M^2_{\Delta_i}-i M_{\Delta_i}\Gamma_{\Delta_i}$, with $\Gamma_{\Delta_i}$ the decay width of $\Delta_i$.
The first term in Eq.(\ref{eq:cpasymfac}) is the leading order contribution where we have
\begin{align}
    \mathrm{Im}\left[\frac{1}{\mathcal{S}_1 \mathcal{S}_2^{\ast} }\right] & =
        \frac{
            \left(s - M_{\Delta_1}^2\right) M_{\Delta_2} \Gamma_{\Delta_2}
            - \left(s - M_{\Delta_2}^2\right) M_{\Delta_1} \Gamma_{\Delta_1}
        }{
            \left[\left(s - M_{\Delta_1}^2\right)^2 + M_{\Delta_1}^2\Gamma_{\Delta_1}^2\right]
            \left[\left(s - M_{\Delta_2}^2\right)^2 + M_{\Delta_2}^2\Gamma_{\Delta_2}^2\right]
        }\,.
\end{align}
Wave functions for incoming and outgoing particles are contained in $\mathcal{W}$.
The second and third terms in Eq.~(\ref{eq:cpasymfac}) are interference terms between tree-level and
one-loop amplitudes of the scattering processes. We note that one-loop vertex corrections vanish.
It is easy to see explicitly in Eq.~\eqref{eq:cpasymfac}
why we need to have a complex $\xi$.
Similarly, one can see that if mediators had the same masses and decay widths,
asymmetry would vanish.
It is important to point out here that the scatterings shown in Fig.~\ref{fig:schan}
contain dark sector scalars, including the DM candidate,
indicating that the evolution of asymmetry in leptogenesis
will be linked to the evolution of DM.
Further discussion about unitarity and CPT in scatterings will be given in Sec.~\ref{sec:uniCPT}.

To work out scatterings for the diagrams in Fig.~\ref{fig:schan},
we proceed to diagonalize the matrix of Eq.~\eqref{eq:denmat}.
Note that in general this matrix is not Hermitian,
therefore, diagonalization proceeds via two matrices such that $M_{D}^2 = L^\dag M_{Dd}^2 R$,
where $L$ and $R$ are 2-by-2 matrices
that define new eigenstates $(\Delta_{R1}, \Delta_{R2})^T = R (\Delta_1, \Delta_2)^T$ and
$(\Delta_{L1}^*, \Delta_{L2}^*)^T = L^* (\Delta_1^*, \Delta_2^*)^T$.
It is possible to write the eigenstates in simple forms by neglecting terms of order $[C_{ij}/(M_{\Delta_1}^2 - M_{\Delta_2}^2)]^2$
obtaining~\cite{Hambye:2000ui}
\begin{align}
    \label{eq:DeltaR1}
    \Delta_{R1} & = \Delta_1 - \frac{i C_{12}\Delta_2}{M_{\Delta_1}^2 - M_{\Delta_2}^2}\,, \\
    \Delta_{R2} & = \frac{i C_{12}^*\Delta_1}{M_{\Delta_1}^2 - M_{\Delta_2}^2} + \Delta_2\,,\\
    \Delta_{L1}^* & = \Delta_1^* - \frac{i C_{12}^*\Delta_2^*}{M_{\Delta_1}^2 - M_{\Delta_2}^2}\,, \\
    \Delta_{L2}^* & = \frac{i C_{12}\Delta_1^*}{M_{\Delta_1}^2 - M_{\Delta_2}^2} + \Delta_2^* \,.
\end{align}
This presents a simplified form of the process of diagonalization of the propagators
of the processes mediated by $\Delta_1$ and $\Delta_2$.
The states actually going through the diagonalization process are those contained
in the triplets since they are the physical mediators in the diagrams of Fig.~\ref{fig:schan}.
In the expressions above we can identify the following states:
\begin{align}
    \Delta_{Rn} & \to \delta^0_{Rn}\,,\, \delta^+_{Rn}\,,\, \delta^{++}_{Rn}\,, \\
    \label{eq:DeltaLnStates}
    \Delta_{Ln}^* & \to \delta^{0*}_{Ln}\,,\, \delta^-_{Ln}\,,\, \delta^{--}_{Ln}\,,
\end{align}
with $n=1,2$.
Additionally, the couplings contained in Eq.~\eqref{eq:cab} should correspond to the couplings
to the physical final and initial states that are involved in the scatterings shown in Fig.~\ref{fig:schan}.
In our numerical analysis we consider this details when calculating the scatterings involved in leptogenesis and DM evolution.

\subsection{Lepton asymmetry and dark matter}
\label{sec:asymmdm}

The $2 \to 2$ scatterings relevant to leptogenesis
can be written explicitly as follows
\begin{align}
    \label{eq:2phi02nu}
\Phi_2^0+\Phi_2^0 \to \delta^0_{Rn} \to \nu_j + \nu_k\,,
\quad \Phi_2^{0*}+\Phi_2^{0*} \to \delta^{0*}_{Ln} \to \nu_j + \nu_k\,, \\
    \label{eq:phimphi0lnu}
\Phi_2^{+} + \Phi_2^0 \to \delta^{+}_{Rn} \to \ell_j^{+} + \nu_k\,,
\quad \Phi_2^{-} + \Phi_2^{0*} \to \delta^{-}_{Ln} \to \ell_j^{-} + \nu_k\,, \\
    \label{eq:2phim2l}
\Phi_2^{+} + \Phi_2^{+} \to \delta^{++}_{Rn} \to \ell_j^{+} + \ell_k^{+}\,,
\quad \Phi_2^{-} + \Phi_2^{-} \to \delta^{--}_{Ln} \to \ell_j^{-} + \ell_k^{-}\,.
\end{align}
We used $\Phi_2^0$ as a shorthand for the neutral fields $H^0$ and $A^0$ in $\Phi_2$.
As it was noted in Ref.~\cite{Kang:2023iur},
These scatterings have a negligible contribution to DM evolution
when the DM candidate has a mass in the TeV scale or above,
where DM mostly annihilates to $W^\pm$ pair.
Other contributions to dark matter annihilation
include final states of SM fermion pairs and vector boson pairs $W^\pm$ and Z. 
The evolution of DM and lepton asymmetry can be described solving the following Boltzmann equations
\begin{align}
        \label{eq:YPhi2eq}
    \frac{d Y_{\Phi_2}}{dx} = {} &
        \frac{-s}{H(x) x} \left(Y_{\Phi_2}^2 - Y_{\mathrm{eq},\Phi_2}^2\right)\langle \sigma v\rangle \left(\Phi_2\Phi_2 \to \mathrm{SM\,SM}\right)\,,\\
    \frac{d Y_{\Delta L}}{dx} = {} & \frac{s}{H(x) x} \bigl[
            \left(Y_{\Phi_2}^2 - Y_{\mathrm{eq},\Phi_2}^2\right)\langle \sigma v\rangle_\delta \left(\Phi_2\Phi_2 \to LL\right) \nonumber\\
        & - 2 Y_{\Delta L} Y_{\mathrm{eq},\Phi_2}^2 Y_{\mathrm{eq},\ell}^{-1}\langle \sigma v\rangle_\mathrm{tot} \left(\Phi_2\Phi_2 \to LL\right) \nonumber\\
        & - 2 Y_{\Delta L} Y_{\mathrm{eq},\Phi_2} \langle \sigma v\rangle_\mathrm{tot} \left(\Phi_2 \bar{L}\to\Phi_2^* L\right) \bigr]
    \label{eq:ydel}
\end{align}
where we have defined, as usual, $Y_{\_\_} = n_{\_\_}/s$
the number density divided by entropy density,
and $x=m_\mathrm{L\Phi_2}/T$ using $m_\mathrm{L\Phi_2}$
as a placeholder for the mass of the lightest state
of $\Phi_2$.
Note that all the coannihilations expressed in Eqs.~\eqref{eq:2phi02nu} to~\eqref{eq:2phim2l} are used in Eqs.~\eqref{eq:YPhi2eq} and~\eqref{eq:ydel}
with $\Phi_2$ used for components the inert scalar doublet.
The thermally averaged cross sections times velocity follow the typical notation
$\langle \sigma v \rangle$ and are defined as follows
\begin{align}
    \label{eq:sigvdelta}
    \langle \sigma v\rangle_\delta \left(\Phi_2\Phi_2 \to LL\right) & \equiv
    \langle \sigma v\rangle \left(\Phi_2\Phi_2 \to LL\right)
    - \langle \sigma v\rangle \left(\Phi_2^*\Phi_2^* \to \bar{L}\bar{L}\right)\,, \\
    \label{eq:sigvtot}
    \langle \sigma v\rangle_\mathrm{tot} \left(\Phi_2\Phi_2 \to LL\right) & \equiv
    \langle \sigma v\rangle \left(\Phi_2\Phi_2 \to LL\right)
    + \langle \sigma v\rangle \left(\Phi_2^*\Phi_2^* \to \bar{L}\bar{L}\right)\,.
\end{align}
For the Hubble parameter we use
\begin{equation}
    H(x) = \sqrt{\frac{8 \pi^3 g_*(T)}{90}} \frac{m_\mathrm{L\Phi_2}^2}{x^2 M_\mathrm{Pl}} \, .
\end{equation}
For the last term of Eq.~\eqref{eq:ydel},
the processes that are counted in $\Phi_2 \bar{L}\to\Phi_2^* L$ are
\begin{align}
    \Phi_2^0 + \bar{\nu}_j \to \Phi_2^{0*} + \nu_k\,, \\
    \Phi_2^- + \bar{\ell}_j \to \Phi_2^{0*} + \nu_k\,, \\
    \Phi_2^- + \bar{\nu}_j \to \Phi_2^{0*} + \ell_k\,, \\
    \Phi_2^- + \bar{\ell}_j \to \Phi_2^+ + \ell_k\,.
\end{align}

We obtain the evolution of DM number density
and of the leptonic asymmetry, $Y_{\Delta L}$,
by solving Eqs.~\eqref{eq:YPhi2eq} and~\eqref{eq:ydel}.
Then, we use the standard electroweak sphaleron process~\cite{Kuzmin:1985mm}
to convert leptonic asymmetry to baryonic asymmetry, $Y_{\Delta B}$,
by using the expression $Y_{\Delta B} = -(28/51)Y_{\Delta L}$~\cite{Harvey:1990qw}
calculated at the sphaleron temperature of $T_\mathrm{sph} = 131.7 \pm 2.3$~GeV~\cite{DOnofrio:2014rug}.
We require that the processes responsible for washout
freeze-out before the DM does,
this results in a requirement that the lightest state in $\Phi_2$
has a mass above $\mathcal{O}(0.1)$~TeV.
This means that the main annihilation channel controlling DM freeze-out
is to $W^+ W^-$,
leaving annihilations into leptons as subleading
and most likely freezing-out before DM.
It is well known that in this mass range,
the components of $\Phi_2$ have nearly degenerate masses
if one wants to have the correct relic density,
thus requiring all coannihilations to be included.
We will use the currently measured relic density $\Omega h^2 = 0.120\pm 0.001$
and the baryon number asymmetry $Y_{\Delta B} = (8.718 \pm 0.004)\times 10^{-11}$
as reported by the Planck collaboration~\cite{Planck:2018vyg}.

\subsection{Unitarity and CPT}
\label{sec:uniCPT}

Here we make a small comment on unitarity and CPT
in this work.
It is known that $CP$-violation from $2 \to 2$ scatterings is constrained
by unitarity and CPT~\cite{Pilaftsis:1997dr,Roulet:1997xa,Pilaftsis:1998pd}.
From unitarity we have the requirement that all the scatterings
with outgoing or incoming states $L_l L_m$ meet the following condition
\begin{equation}
    \sum_j |\mathcal{M}(j \to L_l L_m)|^2 = \sum_j |\mathcal{M}(L_l L_m \to j)|^2
\end{equation}
where we have used $j$ to represent the incoming state on the left-hand side
and the outgoing state in the right-hand side of this equation.
Those states have to be summed over.
The different flavors are indicated by $l$ and $m$.
From Eqs.~\eqref{eq:2phi02nu} to~\eqref{eq:2phim2l}
we know the contributions from $\Phi_2\Phi_2$ to incoming and outgoing states,
so we separate them from the sum as follows
\begin{align}
   |\mathcal{M} & (\Phi_2\Phi_2 \to L_l L_m)|^2 + \sum_{j\neq \Phi_2 \Phi_2} |\mathcal{M}(j \to L_l L_m)|^2 \nonumber\\
   & = |\mathcal{M}(L_l L_m \to \Phi_2\Phi_2)|^2 + \sum_{j\neq \Phi_2 \Phi_2} |\mathcal{M}(L_l L_m \to j)|^2\,.
   \label{eq:uniphi2sep}
\end{align}
The states that remain on the sum include the components of $\Phi_1$, the triplets $\Delta_n$
and pairs of leptons.
From applying CPT invariance on the left-hand side and moving all the sums to the right-hand side
we obtain
\begin{align}
    |\mathcal{M} & (\Phi_2\Phi_2 \to L_l L_m)|^2 - |\mathcal{M}(\Phi_2^*\Phi_2^* \to \bar{L}_l \bar{L}_m)|^2   \nonumber\\
    & =
    \sum_{j\neq \Phi_2 \Phi_2} |\mathcal{M}(j_\text{CPT} \to \bar{L}_l \bar{L}_m)|^2 - \sum_{j\neq \Phi_2 \Phi_2} |\mathcal{M}(j \to L_l L_m)|^2\,.
    \label{eq:uniCPT}
\end{align}
On the left-hand side of this equation we have the processes of Eqs.~\eqref{eq:2phi02nu} to~\eqref{eq:2phim2l}
and all the other processes on the other side.
Note that processes where $j = L_l L_m$
cancel between sums on the right-hand side.
Both sides of this equation contain scatterings mediated by $\Delta_1$ and $\Delta_2$,
therefore, they are subject to the same phases and mixing described
in the first part of Sec.~\ref{sec:asymmetry}.
This results in both sides being non-zero.
The processes contained in the left-hand-side of Eq.~\eqref{eq:uniCPT}
are relevant for the evolution of asymmetry obtained from Eq.~\eqref{eq:ydel}
when solving the Boltzmann equations.

It is worthwhile to emphasize that the vanishing of CP asymmetries derived in Refs.~\cite{Pilaftsis:1997dr, Roulet:1997xa, Pilaftsis:1998pd} applies to the sum over all possible final states for a given initial state, as enforced by unitarity and CPT invariance.
In contrast, the lepton asymmetry we compute corresponds specifically to the subset of channels that violate lepton number, such as $\Phi_2 \Phi_2 \to LL$, whose CP-odd contributions need not cancel once separated from lepton-number–conserving processes.
While the total CP asymmetry summed over all channels vanishes, the projection onto lepton-number–violating channels can be nonzero, and it is precisely this component that enters the Boltzmann equations and is subsequently converted into baryon asymmetry via sphaleron transitions.
Therefore, our calculation of lepton asymmetry does not contradict the cancellation arguments of Refs.~\cite{Pilaftsis:1997dr, Roulet:1997xa, Pilaftsis:1998pd}.

\subsection{Numerical results}
\label{sec:DMnumbers}

\begin{table}[tb]
    \center
    \setlength\tabcolsep{0.2cm}
    \begin{tabular}{lcccc}
        \toprule
        Parameter                  &   BP1    &   BP2  \\
        \cmidrule(r){1-1} \cmidrule(lr){2-3}
        $u$ [$10^{-10}$~GeV]       &  1.502 & 1.065\\
        $\tan\beta$                &  4.768 & 4.765\\
        $\mu_{11}$ [$10^{-8}$~GeV] &  3.0 & 4.780\\
        $\mu_{21}$ [$10^{-8}$~GeV] &  6.911 & 5.059\\
        $\mu_{12}$ [$10^{-1}$~GeV] &  1.488 & 1.653\\
        $\mu_{22}$ [$10^{-6}$~GeV] &  3.091 & 5.424\\
        $|\xi|$ [$10^{-2}$]        &  1.140 & 1.154\\
        $\mathrm{ang}(\xi)$ [rad]  &  $-0.9538$ & $-0.9549$\\
        $\phi_1$ [rad]             &  $-2.386$ & $-2.728$\\
        $\phi_2$ [rad]             &  $-0.1452$ & 0.2269\\
        $m_{H^0}$ [TeV]            &  \multicolumn{2}{c}{2.0}\\
        $m_{A^0}$ [TeV]            &  \multicolumn{2}{c}{2.001}\\
        $M_{\Phi_2^\pm}$ [TeV]     &  \multicolumn{2}{c}{2.008}\\
        $\lambda_A$                &  \multicolumn{2}{c}{$10^{-3}$}\\
        $m_{\nu_1}$ [eV]           &  \multicolumn{2}{c}{$10^{-5}$}\\
        $M_{12}^2$ [GeV$^2$]       &  \multicolumn{2}{c}{$(10^{-6})^2$}\\
        \bottomrule
    \end{tabular}
    \caption{\label{tab:benchmarkpts}%
        Two benchmark points, labeled BP1 and BP2,
        for which we obtain suitable DM relic density
        and baryon asymmetry.
        These are consistent with the evolutions 
        for $\Omega h^2$ and $Y_{\Delta B} = -(28/51) Y_{\Delta L}$
        shown in Fig.~\ref{fig:BPevolution}.
        The neutrino eigenstate $\nu_1$ is assumed to be the lightest,
        i.e., neutrinos have normal mass hierarchy.
        We used $\lambda_A = \lambda_{\Phi 12}  + \lambda'_{\Phi 12} - \lambda_5$.
    }
\end{table}

\begin{table}[tb]
    \center
    \setlength\tabcolsep{0.2cm}
    \begin{tabular}{lcccc}
        \toprule
                  & BP1 & BP2 \\
        \cmidrule(r){1-1} \cmidrule(lr){2-3}
         $Y^\Delta_{(1, 1)}$    & $-0.00491 + i 0.04129$ & $-0.02237 + i 0.02083$\\
         $Y^\Delta_{(1, 2)}$    & $-0.1792 + i 0.0243$ & $-0.2644 + i 0.0720$\\
         $Y^\Delta_{(1, 3)}$    & $-0.05340 - i 0.05526$ & $-0.05757 + i 0.00488$\\
         $Y^\Delta_{(2, 2)}$    & $0.5565 + i 0.1693$ & $0.7666 - i 0.1177$\\
         $Y^\Delta_{(2, 3)}$    & $0.5903 + i 0.0634$ & $0.8478 - i 0.1703$\\
         $Y^\Delta_{(3, 3)}$    & $0.4360 + i 0.1376$ & $0.5987 - i 0.0871$\\
        $M_{\Delta_1}$ [GeV] & 6457 & 9679 \\
        $M_{\Delta_2}$ [GeV] & 4488 & 4562 \\
        $\sigma_\mathrm{SI}$ [cm$^2$] & \multicolumn{2}{c}{$8.911\times 10^{-51}$} \\
        $\langle \sigma v (W^+ W^-) \rangle$ [cm$^3$ s$^{-1}$] & \multicolumn{2}{c}{$3.67\times 10^{-26}$} \\
        \bottomrule
    \end{tabular}
    \caption{\label{tab:benchmarkptsres}%
        Numerical results corresponding to the inputs
        of Table~\ref{tab:benchmarkpts}.
    }
\end{table}

\begin{figure}[tb]
    \center
    \includegraphics{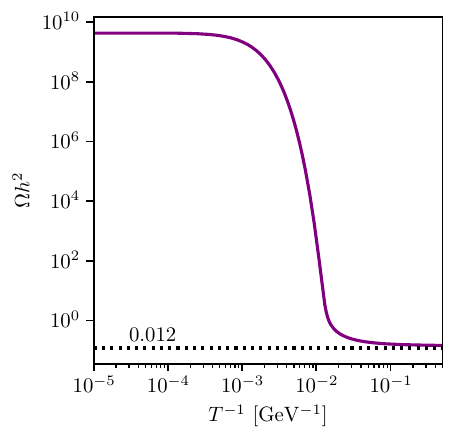}
    \includegraphics{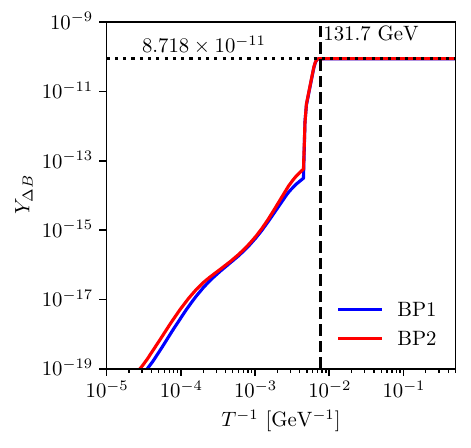}
    \caption{\label{fig:BPevolution}%
        Left: Resulting evolution for $\Omega h^2$ as a function of $T^{-1}$,
        obtained from solving Eqs.~\eqref{eq:YPhi2eq} and~\eqref{eq:ydel}
        with the inputs from Table~\ref{tab:benchmarkpts}.
        Since all the parameters relevant for DM evolution
        are the same in both benchmark points (effect of annihilation into SM fermions is negligible), 
        their solutions follow the same line.
        The horizontal dotted line shows the measured relic density.
        Right: Evolution for $Y_{\Delta B} = -(28/51) Y_{\Delta L}$ as a function of $T^{-1}$
        from solving the same equations and for the same input benchmark points.
        The central value for $Y_{\Delta B} = (8.718 \pm 0.004)\times 10^{-11}$
        is shown as a horizontal dotted line,
        while the dashed vertical line corresponds to the sphaleron temperature,
        $T_\mathrm{sph} = 131.7 \pm 2.3$~GeV, in the $T^{-1}$ axis.
    }
\end{figure}

To obtain evolutions for DM and leptonic asymmetry,
we begin by implementing our model,
presented in Secs.~\ref{sec:tripidm} and~\ref{sec:tripneu},
into \texttt{CalcHEP}~\cite{Belyaev:2012qa}
to obtain the necessary squared amplitudes and decay widths.
From these we calculate averaged cross sections
that are needed to solve Eqs.~\eqref{eq:YPhi2eq} and~\eqref{eq:ydel}.
Additionally, we have taken care of the diagonalization of $M_D^2$
and the corresponding propagator eigenstates and eigenvalues
to include the effects discussed in Sec.~\ref{sec:asymmetry}
into the solution of the Boltzmann equations.
Note that Eqs.~\eqref{eq:DeltaR1} to~\eqref{eq:DeltaLnStates}
need the condition that
$[C_{ij}/(M_{\Delta_1}^2 - M_{\Delta_2}^2)]^2$
is negligible.
Although we always use numerical eigenvalues and eigenstates of $M_D^2$,
during our calculation we make sure that this condition is satisfied,
at least up to scattering energies with relevant contributions
after Boltzmann suppression.
While the number of parameters in the model is quite large,
only a few parameters are expected to have a sizable
effect on relic density evolution and baryon asymmetry.
Take the parameter $u$, for example, when we assume that Yukawa couplings
do not set the neutrino mass scale, the smallness of $u$ takes that role.
We expect the parameters $\mu_{11}$, $\mu_{21}$, and $\beta$
to have an important place due to their involvement in triplet masses,
considering that $u$ is already used to fix neutrino mass scale.
Another parameter that could be relevant to triplet masses is $M^2_{12}$,
however, being part of the soft symmetry breaking part of the scalar potential,
we fix it to a small value that simplifies most expressions related to triplet mixing.
The two parameters $\mu_{12}$, $\mu_{22}$, and $\xi$
appear explicitly in $M_D^2$ and in Eq.~\eqref{eq:cpasymfac}
and are expected to be relevant
for the evolution of leptonic asymmetry.
In the leptonic sector, we used the central values for the measured mixing angles, $CP$ phase
and the squared mass differences in the right-hand side of Eq.~\eqref{eq:yukawa_upmns}.
We take the measured values from the global fit performed by NuFIT~(6.0)~\cite{Esteban:2024eli,nufitwebsite}.
We also fix the value of the mass of $\nu_1$,
which we take to be the lightest neutrino mass eigenstate (normal hierarchy).
This leaves $\phi_1$ and $\phi_2$ as two more free parameters.
Considering that we assume DM at the TeV order,
annihilations are dominated by $W^\pm$ pair final state,
with annihilation into leptons being negligible.
This makes possible to fix the parameters relevant to DM evolution
while varying the parameters relevant to leptonic asymmetry evolution.
The parameters relevant to DM evolution are the masses $m_{H^0}$, $m_{A^0}$ and $M_{\Phi^\pm_2}$,
and the coupling $\lambda_A = \lambda_{\Phi 12} + \lambda'_{\Phi 12} - \lambda_5$.
For all the parameters deemed relevant we show numerical inputs
in Table~\ref{tab:benchmarkpts}
for two benchmark points that realize the correct relic density
and baryon asymmetry.

\begin{figure}[tb]
    \center
    \includegraphics{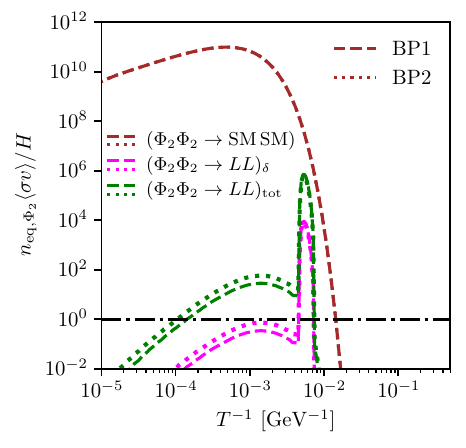}
    \caption{\label{fig:nsigmavh}%
        Interaction rates obtained for the scatterings used in Eqs.~\eqref{eq:YPhi2eq} and~\eqref{eq:ydel},
        divided by the Hubble parameter.
        In the case of $\Phi_2 \Phi_2 \to \mathrm{SM\,SM}$, displayed in brown, it is dominated by annihilations into $W^\pm$ and the lines
        for both benchmark points are the same.
        For the other scatterings BP1 is shown as dashed lines while BP2 is shown as dotted lines.
        The difference $(\Phi_2 \Phi_2 \to L L)_\delta$ (Eq.~\eqref{eq:sigvdelta}) is shown in magenta
        and the sum $(\Phi_2 \Phi_2 \to L L)_\mathrm{tot}$ (Eq.~\eqref{eq:sigvtot}) is shown in green.
    }
\end{figure}

In Table~\ref{tab:benchmarkpts},
the first noticeable detail is the smallness of $u$.
This is explained by the need of it setting the neutrino mass scale,
in combination with the value of the Yukawa couplings.
Yukawa couplings are not allowed to be too small,
combining this with the assumed mass for the lightest neutrino mass eigenstate,
we obtain $u$ of $\mathcal{O}(10^{-10})$~GeV.
In the case of $\tan\beta$, we see that it does not deviate much in the two benchmark points,
and has a value that does not represent a strong hierarchy among the VEVs of the triplets.
In the case of $\mu_{11}$ and $\mu_{21}$,
we see that they are of the same order ($\mathcal{O}(10^{-8})$)~GeV,
mostly due to their relationship to $u$ through minimization of the potential,
and the requirement of TeV masses for the triplets.
The situation is different for $\mu_{12}$ and $\mu_{22}$,
which are of $\mathcal{O}(10^{-1})$~GeV and $\mathcal{O}(10^{-6})$~GeV,
respectively.
Together with the size of $\xi$, this creates a hierarchy among the values of the $M_D^2$ matrix,
that is reflected into asymmetry evolution when solving the Boltzmann equations.
The corresponding numerical results for the benchmark points are displayed in Table~\ref{tab:benchmarkptsres},
where we can see that the Yukawa couplings are not disproportionately small.
The masses for the triplets tend to be in the upper TeV order, almost reaching 10~TeV
in one case.
On the right side of Fig.~\ref{fig:BPevolution},
we show the resulting DM relic density evolution.
As commented before, for quantities relevant to this evolution we use the same
values for both benchmark points so they follow the same evolution.
We see that DM starts freezing out at around $T^{-1} \sim 10^{-2}$~GeV$^{-1}$,
something we can expect after looking at the brown dashed line in Fig.~\ref{fig:nsigmavh}
which crosses the 1 horizontal dash-dotted line at around the same place.
This is convenient because it gives a boost to asymmetry just before reaching
the sphaleron temperature at $\sim 131.7$~GeV,
as can be seen in the left side of Fig.~\ref{fig:BPevolution}.
This boost is due to dark matter deviating from equilibrium
and increasing asymmetry via the first term of Eq.~\eqref{eq:ydel}.
This is in combination with an increased annihilation rate
to leptons before $T^{-1} = 10^{-2}$~GeV$^{-1}$.
This can be understood by the masses of the triplets allowing
resonant annihilations around that temperature.
These details come together into the evolution of asymmetry displayed
in the left side of Fig.~\ref{fig:BPevolution},
where we see first an increase in asymmetry that is considerably boosted
at around $T^{-1} \sim 5\times 10^{-3}$~GeV$^{-1}$.
The shape of the annihilation rates for $(\Phi_2 \Phi_2 \to LL)_{(\delta,\text{tot})}$
in Fig.~\ref{fig:nsigmavh} shows some correspondence to the accelerated growth in asymmetry.

To comment on experimental probes that constraint this model,
we have three in particular
consisting of searches for decays of doubly charged scalars,
direct detection limits for DM
and indirect detection of DM, in our case mostly via annihilations to $W^\pm$ pair.
The searches for decays of doubly charged scalars have been performed at ATLAS and CMS,
putting a lower bound on the mass of this scalars at around 1~TeV~\cite{CMS:2017pet,ATLAS:2017xqs}.
In our case, at least for our benchmark points,
our results shown in Table~\ref{tab:benchmarkptsres} have
the triplets with masses on the TeV scale and above, putting them safely above these searches.
For the case of direct and indirect detection of DM, we use \texttt{micrOMEGAs}~\cite{Belanger:2020gnr}
to obtain spin independent direct detection cross sections, $\sigma_\mathrm{SI}$,
and averaged cross sections times velocity for annihilations to $W^\pm$ pairs, $\langle \sigma v (W^+ W^-) \rangle$,
and report the results in Table~\ref{tab:benchmarkptsres}.
In the case of direct detection, we see that our result for the cross section, $8.911\times 10^{-51}$~cm$^2$,
is several orders of magnitude below the current limit
obtained by the LZ collaboration~\cite{LZ:2024zvo}.
For indirect detection, the most constraining result has been obtained by the H.E.S.S. Collaboration,
putting the limit as low as $3.7\times 10^{-26}$~cm$^3$s$^{-1}$
for DM at a mass of 1.5~TeV~\cite{HESS:2022ygk}.
For the mass of our DM, sitting at $\sim 2$~TeV,
the limit set by H.E.S.S. is closer to the typical $\langle \sigma v \rangle$ required for correct thermal relic density,
with our result of $3.67\times 10^{-26}$~cm$^2$~s$^{-1}$ slightly below this limit.

\subsection{Comparing results for two mechanisms: one-loop against tree-level}

As mentioned in Sec.~\ref{sec:asymmetry},
the approach we followed in this work
considers one-loop contributions to the propagators of the triplets.
These one-loop contributions are responsible for changing between components of $\Delta_1$ and $\Delta_2$
in the middle of the scattering.
This is in contrast to Ref.~\cite{Kang:2023iur} where these contributions were neglected.
One change that we had to consider, is satisfying the condition above Eq.~\eqref{eq:DeltaR1},
something that notably affected the size of the Yukawa couplings.
This required a choice of smaller lightest neutrino mass
resulting in smaller Yukawa couplings.
In this work, the obtained Yukawa couplings are approximately one order of magnitude smaller than in the previous work at tree-level.
Correspondingly, we also obtain $u$, $\mu_{11}$ and $\mu_{21}$ one order of magnitude larger,
in order to keep the triplet masses at TeV level.
One big change is the order of magnitude of $\mu_{22}$, which is five orders of magnitude smaller than $\mu_{12}$,
while in Ref.~\cite{Kang:2023iur} both were at a similar size.
This is related to achieving enough mixing in the matrix of Eq.~\eqref{eq:denmat} to obtain sizable asymmetry.
For the same reason, the size of $\xi$ in this work is two orders of magnitude smaller than in the tree-level case.
When looking specifically at the evolution of asymmetry, we see in Fig.~3 of Ref.~\cite{Kang:2023iur}
that it evolved smoothly due to averaged cross sections
for $(\Phi_2 \Phi_2 \to LL)_{\mathrm{tot},\delta}$ that are well behaved in Fig.~4 of the same work.
While in this work, the same averaged cross sections can be seen to increase sharply around $T\sim 200$~GeV in Fig.~\ref{fig:nsigmavh}.
This results in the accelerated increase in asymmetry than can be seen in Fig.~\ref{fig:BPevolution}.
Expectedly, these notable changes involve mainly parameters that are related to the matrix of Eq.~\eqref{eq:denmat}
and the scatterings that are affected by it, resulting in asymmetry evolving very differently than it did in Ref.~\cite{Kang:2023iur}.

\section{Inflation and reheating}
\label{sec:inflation}

We now turn to the inflationary implications of the model, focusing on the dynamics of the neutral scalar fields with non-minimal gravitational couplings.
A key motivation of this section is to demonstrate how inflation can be embedded within a unified framework that also addresses other unresolved problems of the SM. 
We examine whether those scalar fields can successfully drive inflation consistent with both cosmological observations and phenomenological constraints from dark matter and baryon asymmetry studied above.

For the study of the inflationary dynamics, we only focus on the neutral components of the scalar fields defined as
\begin{equation}
\Phi_1^0=\frac{1}{\sqrt{2}}h_1\,,~~\Phi_2^0=\frac{1}{\sqrt{2}}h_2 e^{i\vartheta}\,,~~
\Delta_1^0=\frac{1}{\sqrt{2}}\delta_1 e^{i\alpha_1}\,,~~\Delta_2^0=\frac{1}{\sqrt{2}}\delta_2 e^{i\alpha_2}\,.
\end{equation}
The Lagrangian terms of neutral components of the scalars in the Jordan framework are given by
\begin{align}
    \frac{\cal L_J}{\sqrt{-g_J}} = & -\frac{1}{2}M^2_{\rm Pl}R+\Big(\xi_1 h_1^2+\xi_2 h_2^2+\xi_{\delta_1}\delta_1^2+\xi_{\delta_2}\delta_2^2 \Big)R \nonumber\\
 &-\left|D_\mu h_1\right|^2-\left|D_\mu h_2\right|^2-\left|D_\mu \delta_1\right|^2-\left|D_\mu \delta_2\right|^2 \nonumber\\
 &-V_J\left(\Phi_1^0,\Phi_2^0,\Delta_n^0 \right) \, .
\end{align} 
The Lagrangian contains the non-minimal couplings, $\xi_i$ ($i=1,2,\delta_1,\delta_2$), that are assumed to be all positive to avoid a potential instability at large field values. In this study, we assume all the parameters to be real.

First, by making a Weyl transformation of the metric $g^J_{\mu\nu}=g^E_{\mu\nu}/\Omega^2$ with
\begin{equation}\label{conformalxform}
\Omega^2\equiv 1+\frac{1}{M^2_{\rm Pl}}\left(\xi_1 h_1^2+\xi_2 h_2^2 +\xi_{\delta_1}\delta_1^2+\xi_{\delta_2}\delta_2^2 \right).
\end{equation}
we obtain the Einstein frame action as
\begin{align}
\label{einsteinaction}
\frac{\calL_E}{\sqrt{-g_E}} =& -\frac{1}{2}M^2_{\rm Pl}R-\frac{3}{4}\Big[M_{\rm Pl}\partial_\mu \log \Omega^2 \Big]^2
-\frac{(\partial_\mu h_1)^2+(\partial_\mu h_2)^2+(\partial_\mu \delta_1)^2+(\partial_\mu \delta_2)^2}{2\Omega^2} \nonumber \\
&-\frac{(h_2\partial_\mu \vartheta )^2+(\delta_1\partial_\mu \alpha_1)^2+(\delta_2 \partial_\mu \alpha_2)^2}{2\Omega^2}
 - V_E(\Phi_1^0,\Phi_2^0,\Delta_n^0) \, ,
\end{align}
where
\begin{equation}
\label{einsteinpotential}
V_E(\Phi_1^0,\Phi_2^0,\Delta_n^0) = \frac{V_J}{\Omega^4} \, .
\end{equation}
Here we have dropped the gauge interactions.

Ignoring the mass terms in the potential,
(\ref{einsteinaction}) and (\ref{einsteinpotential}) with $\phi^I=\{h_1,h_2,\delta_1,\delta_2,\vartheta, \alpha_1,\alpha_2\}$ are rewritten as
\begin{equation}
\frac{{\cal L}_E}{\sqrt{-g_E}}= -\frac{1}{2}M^2_{\rm Pl}R-\frac{1}{2}G_{IJ}\partial_\mu \phi^I\partial^\mu\phi^J-V_E(\phi^I) \, ,
\end{equation}
where 
\begin{equation}
G_{IJ} =\frac{\delta_{IJ}}{\Omega^2}+\frac{3M^2_{\rm Pl}}{2} \frac{\partial  \log \Omega^2 }{\partial \phi^I}\frac{\partial  \log \Omega^2 }{\partial \phi^J}.
\end{equation}

\subsection{Two-Higgs triplet inflation}
We consider an inflation scenario along the direction where $h_i=0$. In this case, we set $\phi^I=\{ \delta_1, \delta_2 \}$ in the Lagrangian ${\calL_E}$.
In this setup, $\Omega^2$ takes the form $1+(\xi_{\delta_1}\delta^2_1+\xi_{\delta_2}\delta^2_2)/M^2_{\rm Pl}$.
In a large field limit, where $\xi_1 h^2_1+\xi_2 h^2_2\gg M^2_{\rm Pl}$,  the mass terms in the Lagrangian can be neglected, leaving only quartic terms in the scalar potential.
By further neglecting the quartic couplings in $V_{\rm SB}$, we obtain
the explicit form of the matrix $G_{IJ}$ and the scalar potential $V_E(\phi^I)$, which are given by
\begin{align}
G_{IJ} = & \frac{1}{\Omega^2}
\begin{pmatrix}
1+\dfrac{6\xi^2_{\delta_1} \delta^2_1}{\Omega^2M^2_{\rm Pl}} & \dfrac{6\xi_{\delta_1}\xi_{\delta_2} \delta_1\delta_2}{M^2_{\rm Pl}}
\\
\dfrac{6\xi_{\delta_1}\xi_{\delta_2} \delta_1\delta_2}{M^2_{\rm Pl}} & 1+\dfrac{6\xi^2_{\delta_2} \delta^2_2}{\Omega^2M^2_{\rm Pl}} 
\end{pmatrix} \, ,
\\
V_E(\phi^I) = & \frac{\lambda_{\Delta_1}\delta^4_1+\lambda_{\Delta_2} \delta^4_2+2 \lambda_{\Delta_M}\delta_1^2\delta_2^2}
{8\Omega^4} \, 
\label{epot2}
\end{align}
where $\lambda_{\Delta_M} \equiv \lambda_{\Delta 12} + \lambda_{\Delta 21} + \lambda'_{\Delta 12} + \lambda'_{\Delta 21}$ comes from the quartic couplings with two $\Delta_1$ and two $\Delta_2$.

Next, we redefine the fields as follows:
\begin{align}
\varphi=  \sqrt{\frac{3}{2}}M_{\rm Pl}\log(\Omega^2) , ~~
s =  \frac{\delta_2}{\delta_1} \, ,  \label{newfield2}
\end{align}
This leads to the action in the form~\cite{Lebedev:2011aq}
\begin{align}
\label{largefieldaction}
\frac{\calL_E}{\sqrt{-g_E}} \approx {}& -\frac{1}{2}M^2_{\rm Pl}R - \frac{1}{2}\left(1+\frac{1}{6}\frac{s^2+1}{\xi_{\delta_2} s^2+\xi_{\delta_1}}\right)(\partial_\mu \varphi)^2-\frac{1}{\sqrt{6}}\frac{(\xi_{\delta_1}-\xi_{\delta_2}) s}{\left(\xi_{\delta_2} s^2+\xi_{\delta_1}\right)^2}(\partial_\mu \varphi)(\partial^\mu \tilde{s})
\nonumber\\
&-\frac{1}{2}\frac{\xi^2_{\delta_2} s^2+\xi^2_{\delta_1}}{\left(\xi_{\delta_2} s^2+\xi_{\delta_1}\right)^3}(\partial_\mu \tilde{s})^2-V_E(\varphi,s) \, ,
\\
\label{potentialwomin}
V_E(\varphi,s)= {}& M^4_{\rm Pl}\frac{\lambda_{\Delta_1}+\lambda_{\Delta_2} s^4+2\lambda_{\Delta_M} s^2}{8\left(\xi_{\delta_2} s^2+\xi_{\delta_1}\right)^2}\,\left(1-e^{-\frac{2\varphi}{\sqrt{6}M_{\rm Pl}}}\right)^2  \, .
\end{align}
where $\tilde{s}=M_{\rm Pl} s$.
The potential given in (\ref{potentialwomin}) belongs to a class of potentials called Starobinsky potentials \cite{Starobinsky:1979ty},
which is almost flat at large field values ensuring slow roll.
To render the kinetic terms in Eq.(\ref{largefieldaction}) canonical, we consider  the scenario where the non-minimal couplings $\xi_{\delta_1}$ and $\xi_{\delta_2}$ are large. Under this assumption,  the kinetic terms at leading order in $1/\xi_{\delta_{1(2)}}$ become
\begin{equation}
    \label{eq:lagkin}
-{\cal L}_{\rm kin}\simeq \frac{1}{2}(\partial_\mu \varphi)^2+\frac{1}{2}\frac{\xi^2_{\delta_2} s^2+\xi^2_{\delta_1}}{\left(\xi_{\delta_2} s^2+\xi_{\delta_1}\right)^3}(\partial_\mu \tilde{s})^2
\end{equation}
We can achieve a canonical form for ${\calL}_{\rm kin}$ by redefining $\tilde{s}$ to $\tilde{s}^{\prime}$, depending on the relative magnitude of
$\xi_{\delta_1}$ and $\xi_{\delta_2}$ as follows:
\begin{align}
    \xi_{\delta_1} \gg {}&  \xi_{\delta_2}, ~~\tilde{s}^{\prime}=\frac{\tilde{s}}{\sqrt{\xi_{\delta_1}}}, \nonumber \\
\xi_{\delta_2} \gg {}&  \xi_{\delta_1}, ~~\tilde{s}^{\prime}=\frac{1}{\sqrt{\xi_{\delta_2}}\tilde{s}}, \nonumber \\
\xi_{\delta_1} = {}&  \xi_{\delta_2}, ~~\tilde{s}^{\prime}=\frac{1}{\sqrt{\xi_{\delta_1}}}\arctan \tilde{s},
\end{align}

\subsection{Single field inflation}
To achieve slow-roll inflation using a single field, we can consider a specific limit of the potential $V_E(\varphi,s)$ in which the field $s$
is stabilized.
By minimizing the scalar potential $V_E(\varphi,s)$ with respect to $s$, we find that three extrema arise:
\begin{equation}
s_0^2=\left\{ \begin{array}{c}
0 \\
\infty \\
\frac{\lambda_{\Delta_1}\xi_{\delta_2}-\lambda_{\Delta_M}\xi_{\delta_1}}{\lambda_{\Delta_2}\xi_{\delta_1}-\lambda_{\Delta_M}\xi_{\delta_2}}
\end{array} \right.
\end{equation}
For a minimum to exist at a finite value of $s_0$,  the following conditions must  be satisfied,
\begin{align}
    \lambda_{\Delta_1}\xi_{\delta_2}-\lambda_{\Delta_M}\xi_{\delta_1}> {}& 0 \, ,
    \label{cond1}\\
    \lambda_{\Delta_2}\xi_{\delta_1}-\lambda_{\Delta_M}\xi_{\delta_2}> {}& 0 \, ,
    \label{cond2} \\
    \lambda_{\Delta_1}\lambda_{\Delta_2}-\lambda_{\Delta_M}^2> {}& 0 \, .
    \label{cond3}
\end{align}
The last condition ensures that no deep minima exist, which could make the electroweak vacuum metastable. It also ensures that the vacuum energy remains positive during inflation, If Eqs.~(\ref{cond1}) and~(\ref{cond2}) are not satisfied, $s_0$ will either be $0$ or $\infty$, This implies that a single neutral scalar, either $\delta_{1}$ or $\delta_2$, drives inflation.
For positive $\lambda_{\Delta_i}$ and $\lambda_{\Delta_M}$, those conditions guarantee the stability of the scalar potential.
Conversely, when $\lambda_{\Delta_M}<0$, Eqs.~(\ref{cond1}) and~(\ref{cond2}) are automatically satisfied, making the third condition (\ref{cond3}) the only remaining  constraint for stability.
For a non-trivial $s_0^2$, inflation is driven by $\varphi$ which represents
a combination of these two scalars $\delta_{i}$. We now focus on this scenario.
When $s=s_0$, the scalar potential simplifies to 
\begin{equation}
V_E(\varphi,s)|_{s=s_0}=\frac{\lambda_{\rm eff}M^4_{\rm Pl}}{4\xi^2_{\rm eff}}\left(1-e^{-2\varphi^{\prime}/\sqrt{6}} \right)^2,
\label{vef}
\end{equation}
where $\varphi^{\prime}=\varphi/M_{\rm Pl}$, $\xi_{\rm eff}=\xi_{\delta_2} s^2_0+\xi_{\delta_1}$
and $\lambda_{\rm eff}=(\lambda_{\Delta_1}+\lambda_{\Delta_2} s^4_0 + \lambda_{\Delta_M}s^2_0)/2$.
By substituting the non-trivial $s^2_0$ into Eq.~(\ref{vef}), we can derive the inflationary vacuum energy as follows
\begin{equation}
V_0=M^4_{\rm Pl}\frac{\lambda_{\Delta_1}\lambda_{\Delta_2}-\lambda^2_{\Delta_M}}{8\left(\lambda_{\Delta_1}\xi^2_{\delta_2}+\lambda_{\Delta_2}\xi^2_{\delta_1}-2\lambda_{\Delta_M}\xi_{\delta_1}\xi_{\delta_2}\right)} \, .
\end{equation}

For most of the parameter space, $\lambda_{\rm eff}$ is not small, so the CMB normalization of density perturbations requires $\xi_{\rm eff}$ to be of $\calO(10^4)$ as will be shown in next sections. Therefore, unitarity is violated at $\mu_U\sim \mpl/\xi_{\rm eff}$, which is well below the Planck scale. Nevertheless, it is possible to maintain the inflationary conditions (\ref{cond1})-
(\ref{cond3}) in the unitarization process of introducing a heavy real scalar as in Higgs portal inflation~\cite{Giudice:2010ka,Lebedev:2011aq}. A detailed treatment of this aspect is beyond the scope of the present work but would not qualitatively alter our inflationary predictions.
We remark that the effective self-coupling of the inflaton, $\lambda_{\rm eff}$, need not be strictly of $\calO(1)$ to satisfy the Higgs mass constraint, unlike the SM Higgs inflation.

To investigate if the inflation scenario we consider can be in consistent with observations, 
we first consider the slow-roll parameters such as  $\epsilon$, $\eta$ and $\zeta$, which  are calculated as
\begin{align}
\epsilon =\frac{1}{2}\left(\frac{1}{V_E}\frac{dV_E}{d\varphi}\right)^2= \frac{4}{3}\, \frac{e^{-4\varphi^{\prime}/\sqrt{6}}}{\left(1-e^{-2\varphi^{\prime}/\sqrt{6}}\right)^2} \, ,
\end{align}
and
\begin{align}
\eta =\frac{1}{V_E}\frac{d^2V_E}{d\varphi^2} = -\frac{4}{3}e^{-2\varphi^{\prime}/\sqrt{6}} \,\frac{1-2e^{-2\varphi^{\prime}/\sqrt{6}}}{\left(1-e^{-2\varphi^{\prime}/\sqrt{6}}\right)^2} \, ,
\end{align}
\begin{equation}
\zeta = \frac{1}{V^2_E}\frac{dV_E}{d\varphi}\frac{d^3V_E}{d \varphi^3}.
\end{equation}

The number of $e$-folds ${\cal N}$ defined as 
\begin{equation}
{\cal N} = \int_e^\star \frac{V_E}{dV_E/d\varphi} d\varphi 
\, ,
\label{efoldint}
\end{equation}
is calculated as
\begin{equation}
    \label{efoldsol}
{\cal N} = \frac{3}{4} \left[ e^{2\varphi_\star^{\prime}/\sqrt{6}} - e^{2\varphi_e^{\prime}/\sqrt{6}} - \frac{2}{\sqrt{6}} \left( \varphi_\star^{\prime} - \varphi_e^{\prime} \right) \right] \, ,
\end{equation}
where the subscripts $\star$ and $e$ denote the moment when the scale of our interest exits the horizon and the end of slow-roll inflation, respectively.

Given that $\epsilon$ should not be greater than one during the slow roll inflation, we determine the value of $ \varphi_e^{\prime}$ as follows:
\begin{align}
    \epsilon(\varphi_e^{\prime})\simeq 1 \Rightarrow~~{\rm  exp}\left(\sqrt{\frac{2}{3}}\varphi_e^{\prime}\right) \simeq {}& 2.15 \nonumber \\
    \varphi_e^{\prime}\simeq {}& 0.94
\label{end-inflation}
\end{align}
For ${\cal N}=60$, using Eq.~\eqref{end-inflation}, we get $\varphi_{\ast}^{\prime}\simeq 5.45$.

Using the result of $\varphi_{\ast}$ with ${\cal N}$ fixed at 60, we determine the tensor to scalar ratio $r$,  the spectral index $n_s$ and  the running of spectral index $n_{rs}$ as follows~\cite{Das:2022qyc}
\begin{align}
r&=16\epsilon \simeq 0.00298, \\
n_s&=1-6\epsilon + 2 \eta \simeq 0.9677, \\
n_{rs}&=-2\zeta -24 \epsilon^2+16\eta \epsilon \simeq  -4.86 \times 10^{-5}
\end{align}

To match the amplitude of the scalar power spectrum~\cite{Bezrukov:2017dyv}, $A_s=V_E/(24 \pi^2 M^4_{\rm Pl} \epsilon)$, with the observational value of $\left(2.101^{+0.031}_{-0.034}\right)\times 10^{-9}$ \cite{Planck:2018vyg}, 
the value of the ratio $\lambda_{\rm eff}/\xi_{\rm eff}^2$ needs to be approximately $3.8\times 10^{-10}$.
Then, $\xi_{\rm eff}$ is related to $\lambda_{\rm eff}$,
\begin{equation}
   \label{eq:xiefflameff}
\xi_{\rm eff}\simeq 5.1\times 10^4 \sqrt{\lambda_{\rm eff}}.
\end{equation}

\subsection{Reheating}
At the end of inflation, the energy density stored in the inflaton field $\varphi$  starts to disperse through
 the annihilation and/or decay into other particles, including those of the SM. This epoch is
 known as the reheating \cite{Allahverdi:2010xz}. It takes the Universe from the matter-dominated phase during
 inflation to the radiation-domination phase. 

As $\varphi$ falls below the Planck scale $M_{\rm Pl}$, the  inflationary potential in Eq.(\ref{vef}) can be approximated as a quadratic potential \cite{Bezrukov:2008ut},
\begin{equation}
V_R=\frac{1}{2} \omega^2 \varphi^2,
\end{equation}
where $\omega^2=\frac{\lambda_{\rm eff}M^2_{\rm Pl}}{3\xi_{\rm eff}^2}$ which suggests the reheating occurs in the harmonic oscillator potential 
well as $\varphi$ undergoes 
coherent oscillations with rapid frequency $\omega$ \cite{Linde:1981mu}.

The equation of motion for inflaton $\varphi$ during reheating can be expressed as,
\begin{equation}
\ddot{\varphi}+3H\dot{\varphi}+V^{\prime}_R=0,
\label{eqofmotion}
\end{equation}
where the dots denote derivatives with respect to the cosmic time $t$, and the prime in $V_R$ stands for the derivative with respect to $\varphi$, and $H$ is the Hubble expansion rate.
In the limit $\omega \gg H$, we get the solution of eq.(\ref{eqofmotion}) as,
\begin{equation}
\varphi=\varphi_0(t)\cos(\omega t),
\end{equation}
where $\varphi_0(t)=\sqrt{\frac{8\xi_{\rm eff}^2}{\lambda_{\rm eff}}}\frac{1}{t}$.

We now define $t_{\rm cr}=\frac{2\xi_{\rm eff}}{\omega}$ as the time at which  the amplitude of $\varphi$ crosses $\varphi_{\rm cr}=\sqrt{\frac{2}{3}}\frac{\mpl}{\xi_{\rm eff}}$, and the quadratic phase ends.
In this scenario, rehating occurs by the productions of  the SM particles through the decay or scattering of the inflaton fields.
The inflaton can decay into the $W$ and $Z$ bosons through the effective coupling terms such as $\frac{g^2}{4\sqrt{6}}\frac{M_{\rm Pl}}{\xi_{\rm eff}}\varphi W^2$ with the weak gauge coupling $g$ and
into the SM Higgs and inert scalar through the coupling terms such as $\lambda_{i}\sqrt{\frac{2}{3}}\frac{M_{\rm Pl}}{\xi_{\rm eff}}\varphi \Phi_j^2$
where $\lambda_i$ denotes the quartic couplings in the scalar potential and $j=1,2$ corresponding to the SM Higgs and inert scalar.
Although those particles do not have a physical mass at the time of reheating, an effective mass arises due to the couplings to inflaton and its oscillations.
For $\omega \gg H$, the explicit expressions of the effective masses are given as \cite{Bezrukov:2008ut}
\begin{align}
m^2_W(\varphi) &= \frac{g^2}{2\sqrt{6}}\frac{M_{\rm Pl} |\varphi|}{\xi_{\rm eff}}, \\
m^2_{h_i}(\varphi) &= \lambda_i \sqrt\frac{2}{3}\frac{M_{\rm Pl} |\varphi|}{\xi_{\delta_{\rm eff}}}, \\
m_F(\varphi) &= y_F \sqrt{\frac{M_{\rm Pl}|\varphi|}{\sqrt{6}\xi_{\rm eff}}}
\end{align}
where $i=1,2$ represents SM Higgs and inert scalar, and $F=t,\nu$ denotes top quark and neutrinos.
We note that the coupling constant $Y_{\nu}$ is of order one in this scenario.
Since they are non-relativistic, due to large couplings, the matter-radiation transition happens only when
the relativistic secondary particles are produced via decays or scatterings of the heavy particles.
When  the number densities of the produced bosons are low, decays of bosons to SM fermions is the dominant channel to produce relativistic particles.
But, if the number densities of the bosons are large, parametric resonance production will be possible.
In the latter case, the dominant channel for producing relativistic particles is the annihilation of $W$ bosons.
For the case of Higgs, the production of the relativistic particles is possible only through decays.
As studied in Ref.~\cite{Bezrukov:2008ut}, the production rate of Higgs is small and its decay to fermions is at a much lower rate
 than the annihilation of $W$ bosons.
Following Ref.~\cite{Choubey:2017hsq},  the radiation density $\rho_{\rm rad}$ is approximately given in our case by
\begin{equation}
\rho_{\rm rad}\simeq \frac{1.46\times 10^{57}}{\sqrt{\lambda_{\rm eff}}}\,.
\label{rho_rad}
\end{equation}
Using this result, we can estimate the reheating temperature $T_{\rm reh}$ from the relation,
\begin{equation}
    \label{eq:rho_rad_Treh}
\rho_{\rm rad}=\frac{\pi^2}{30} g_{*} T^4_{\rm reh},
\end{equation}
where $g_{\ast}$ is the effective number of relativistic degree of freedom at the time of reheating.
Using Eq.(\ref{rho_rad}) and $g_{\ast}=100$, we obtain the reheating temperature $T_{\rm reh}$,
\begin{equation}
    \label{eq:Treh}
T_{\rm reh}\sim 10^{14} ~~\mbox{GeV}.
\end{equation}
We remark that the reheating temperature obtained in our scenario is relatively high compared to typical values in scalar–inflation models. Such a high scale is not inconsistent with the cosmological history adopted here, but in ultraviolet (UV) completions (e.g., supersymmetric extensions) it may raise concerns about the overproduction of unwanted relics such as gravitinos or monopoles. Since our framework is a purely scalar extension of the SM, these issues are not directly relevant, although embedding into more complete theories could potentially impose additional constraints.

\begin{figure}[tb]
    \centering
    \includegraphics{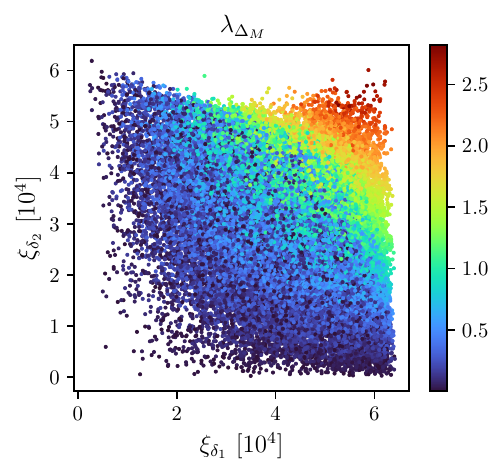}%
    \includegraphics{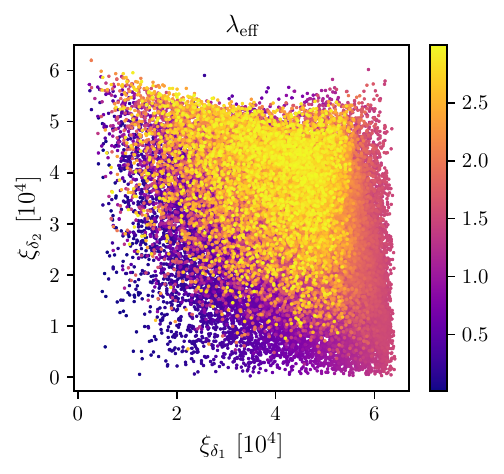}
    \caption{\label{fig:xid1xid2}%
        Distribution of non-minimal couplings $\xi_{\delta_1}$ and $\xi_{\delta_2}$
        that are allowed by the observational value of the amplitude of the scalar power spectrum.
        On the left we show in color how $\lambda_{\Delta M}$ is related to the sizes
        of the non-minimal couplings.
        On the right we do the same but for $\lambda_\mathrm{eff}$.
    }
\end{figure}

\begin{figure}[tb]
    \centering
    \includegraphics{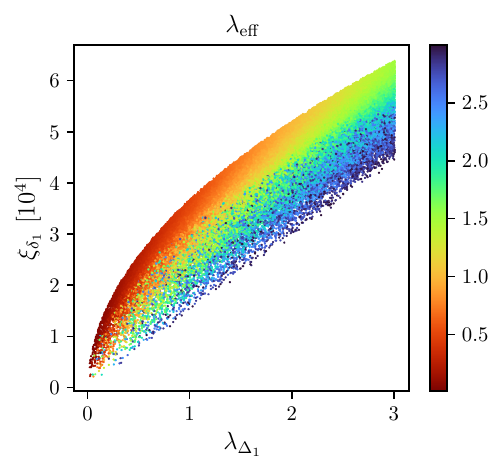}%
    \includegraphics{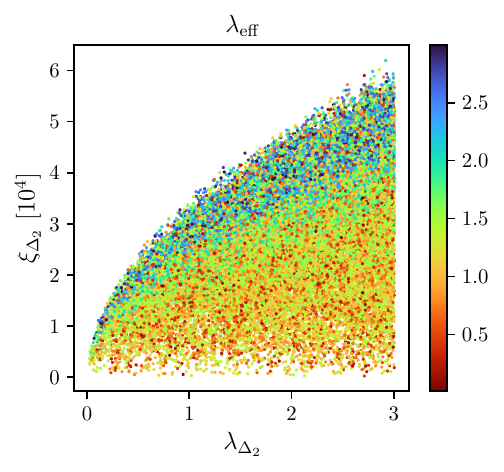}
    \caption{\label{fig:ld1ld2}%
        Relationship between the couplings $\lambda_{\Delta_1}$ and $\lambda_{\Delta_2}$
        and the non-minimal couplings $\xi_{\delta_1}$ and $\xi_{\delta_2}$.
        These points correspond to the points of Fig.~\ref{fig:xid1xid2}
        and are allowed by the observational value of the amplitude of the scalar power spectrum.
        On the left we show the relationship between allowed values for $\lambda_{\Delta_1}$
        and $\xi_{\delta_1}$, with color showing the value for $\lambda_\mathrm{eff}$.
        On the right we do the same for $\lambda_{\Delta_2}$ on the $x$-axis
        and $\xi_{\delta_2}$ on the $y$-axis.
    }
\end{figure}

\begin{figure}[tb]
    \centering
    \includegraphics{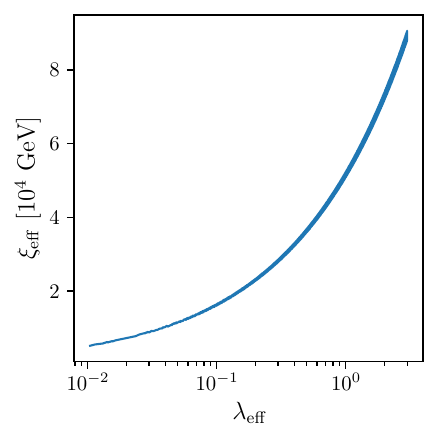}
    \includegraphics{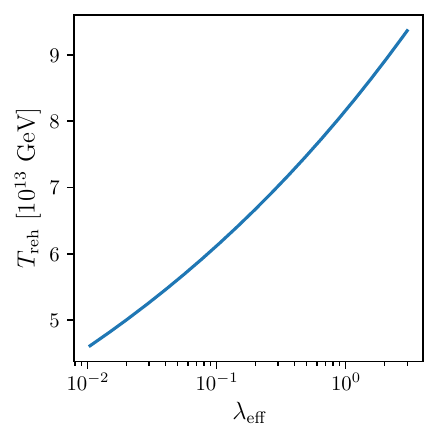}
    \caption{\label{fig:leffTreh}%
        Figures for $\xi_\mathrm{eff}$ (left) and the reheating temperature (right), $T_\mathrm{reh}$, as functions of $\lambda_\mathrm{eff}$.
        The figure for $\xi_\mathrm{eff}$ shows a band due to the $2 \sigma$ range of $A_s$ used to constrain the parameters.
        The line for $T_\mathrm{reh}$ follows the values obtained from relating Eqs.~\eqref{rho_rad} and~\eqref{eq:rho_rad_Treh}.
        The range displayed for $\lambda_\mathrm{eff}$ is consistent with the values obtained
        from the points shown in Figs.~\ref{fig:xid1xid2} and~\ref{fig:ld1ld2}.
    }
\end{figure}

To show the viable parameter space for the couplings
$\lambda_{\Delta_1}$, $\lambda_{\Delta_2}$, $\lambda_{\Delta_M}$,
and the non-minimal couplings $\xi_{\delta_1}$ and $\xi_{\delta_2}$,
we randomly sample their values,
with the couplings of the scalar potential in the range [0,3]
and the non-minimal couplings in the range [100,70\,000].
We check that the conditions of Eqs.~\eqref{cond1} to~\eqref{cond3} are met
and make sure that the prediction for the amplitude of the power spectrum, $A_s$,
is within the $\pm 2\sigma$ according to the value cited above Eq.~\eqref{eq:xiefflameff}.
The value for $\varphi '_*$ is obtained from Eq.~\eqref{efoldsol}
and given below Eq.~\eqref{end-inflation}.
On Fig.~\ref{fig:xid1xid2} we show the distribution of $\xi_{\delta_1}$ and $\xi_{\delta_2}$
allowed by these constraints,
along with color for the tendencies of other parameters.
In particular, on the left side we see that
larger $\lambda_{\Delta_M}$ allows to have both larger $\xi_{\delta_1}$ and $\xi_{\delta_2}$.
On the right side we see that $\lambda_\mathrm{eff}$ tends to be $\mathcal{O}(1)$ for
$\xi_{\delta_{1}}$ and $\xi_{\delta_{2}}$ of $\mathcal{O}(10^4)$,
with some thining of the points for smaller values due to the constraints.
In particular, note that larger $\xi_{\delta_1}$ allows for more uniform larger $\lambda_\mathrm{eff}$.
These values are consistent with our requirement that the non-minimal couplings
are large as was mentioned before Eq.~\eqref{eq:lagkin}.
On Fig.~\ref{fig:ld1ld2} we show relationships between the scalar potential couplings
and the non-minimal couplings.
On the left side we see that larger $\lambda_{\Delta_1}$ allows larger $\xi_{\delta_1}$,
although in a limited range.
In color we see that $\lambda_\mathrm{eff}$ tends to be larger for smaller $\xi_{\delta_1}$,
but that it is allowed to take smaller values with smaller $\lambda_{\Delta_1}$.
On the right side we see that $\lambda_{\Delta_2}$ increases the upper limit for $\xi_{\delta_2}$,
without a noticeable effect on its lower bound.
For these two parameters, we see that $\lambda_\mathrm{eff}$ has larger values
for larger $\xi_{\delta_2}$, however its range does not change much with changes in $\lambda_{\Delta_2}$,
indicating that the size of $\lambda_\mathrm{eff}$ is mostly controlled by $\lambda_{\Delta_1}$,
$\xi_{\delta_1}$ and $\xi_{\delta_2}$.
We can finally relate these parameters to reheating,
using Eqs.~\eqref{rho_rad} and~\eqref{eq:rho_rad_Treh}
to obtain $T_\mathrm{reh}$ as a function of $\lambda_\mathrm{eff}$.
We show this dependence in the right side of Fig.~\ref{fig:leffTreh},
using $\lambda_\mathrm{eff}$ in the range obtained from the parameters discussed above.
We see that the obtained reheating temperature is close to $10^{14}$~GeV,
as was estimated in Eq.~\eqref{eq:Treh}.
For reference, we also show the dependence of $\xi_\mathrm{eff}$
on the right side of Fig.~\ref{fig:leffTreh},
with a size consistent with the estimation given in Eq.~\eqref{eq:xiefflameff}.
Note that the line shown for $\xi_\mathrm{eff}$ is actually a band
that was calculated from its definition given below Eq.~\eqref{vef},
using the values of the couplings shown in Figs.~\ref{fig:xid1xid2} and~\ref{fig:ld1ld2},
and meets the same constraints that were used to obtain said figures.

\section{Common origin to all of the above}

In Secs.~\ref{sec:tripidm}, \ref{sec:asymmetry} and~\ref{sec:inflation} we have developed neutrino masses, dark matter, baryon asymmetry,
inflation and reheating, in their respective sections.
We have described the processes that make them possible
and discussed their place in the parameter space of the model presented in Sec.~\ref{sec:tripidm}.
In all these processes, the triplets play crucial roles,
serving as mediators in the generation of neutrino masses via inverse type-II seesaw,
annihilation of dark matter
and production of asymmetry in the leptonic sector that is later converted to baryonic asymmetry.
They also serve as parts of the inflaton by having non-minimal gravitational couplings.
In this way, the presence of the triplets provides an economical connection
between phenomena that remains unexplained inside the SM.
However, there are other components that play important roles.
In the case of neutrinos, we require trilinear couplings between the SM-like Higgs
and one of the triplets ($\Phi_1^T i\sigma^2 \Delta_n^\dag \Phi_1$), parameterized by the $\mu_{n1}$.
For DM, similarly, we require a trilinear coupling to the inert doublet ($\Phi_2^T i\sigma^2 \Delta_n^\dag \Phi_2$),
parameterized by the couplings $\mu_{n2}$,
that allows the triplets to be mediator between annihilations to $W$ boson pair.
In conjunction with the couplings to leptons, parameterized by Yukawa couplings,
these components allow the appearance of asymmetry in the processes involved in the evolution of the number density
of leptons.
Finally, for inflation and reheating,
the most important element is the interplay between the triplets,
particularly for the couplings parameterized by $\lambda_{\Delta_1}$, $\lambda_{\Delta_2}$
and the sum $\lambda_{\Delta_M}$,
as well as the non-minimal couplings $\xi_{\delta_1}$ and $\xi_{\delta_2}$.
Note that the parameters involved in each of these phenomena happen
to have negligible influence on each other,
with inflation and reheating depending mostly on $\lambda_{\Delta_1}$, $\lambda_{\Delta_2}$,
$\lambda_{\Delta_M}$, $\xi_{\delta_1}$ and $\xi_{\delta_2}$,
while dark matter and baryon asymmetry are mostly controlled by the parameters of Table~\ref{tab:benchmarkpts}.
In a summarized way, this is how the setup presented in Sec.~\ref{sec:tripidm}
provides a common origin for the mechanisms involved
in the phenomena described in the past sections.

\section{Conclusion}
\label{sec:conclusion}

In this work we have explored an extension of the Standard Model scalar sector by introducing two $SU(2)$ triplets together with an inert $SU(2)$ doublet, providing a unified framework that simultaneously accounts for neutrino masses, dark matter, baryon asymmetry, and inflation with reheating. The triplet scalars play a central role: they couple to leptons and generate neutrino masses via an inverse type-II seesaw mechanism without requiring unnaturally small Yukawa couplings, they mediate dark matter annihilations (including asymmetric channels into leptons) as well as CP-violating scatterings responsible for leptogenesis, and through non-minimal gravitational couplings they contribute to the inflaton sector. The inert doublet supplies a viable dark matter candidate, with its freeze-out linked to the generation of the lepton asymmetry.
In this way, the same scalar triplets that determine the evolution of dark matter and baryogenesis also develop into the inflaton
ensuring that inflation and reheating emerge as interconnected aspects of the unified scalar framework rather than as separate additions.

Our numerical analysis demonstrates benchmark points where the observed relic density and baryon asymmetry are simultaneously reproduced, with inflationary observables consistent with Planck constraints. For dark matter, the direct detection cross section is several orders of magnitude below the reach of upcoming experiments~\cite{Billard:2021uyg}, while indirect searches, especially annihilations into $W$ pairs, impose strong constraints and are already close to probing the benchmark scenarios. For inflation, the predicted parameters fall well within the ranges reported by the Planck Collaboration, and the reheating dynamics, dominated by decays into $W$ bosons and subsequent fermionic channels, yield a high reheating temperature compatible with standard cosmology. Collider searches for doubly charged scalars provide additional constraints, but the multi-TeV triplet masses considered here remain beyond the current experimental reach~\cite{deMelo:2019asm}.

In summary, these results show that an economical scalar extension of the SM can provide a common origin for neutrino masses, dark matter, baryogenesis, and cosmic inflation. The framework remains predictive and testable, offering multiple avenues for experimental verification in both collider and astrophysical frontiers in the near future.

\section*{Acknowledgments}

The work of SKK is supported by supported by the National Research Foundation of Korea under grant NRF-2023R1A2C100609111.
The work of RR is supported by a KIAS Individual Grant
with number QP094601 at Korea Institute for Advanced Study.
This work uses computational resources supported by the Center for Advanced Computation at Korea Institute for Advanced Study.

\appendix

\section{Extended scalar sector}
\label{app:scalars}

A complete analysis of the scalar sector, including the scalar potential, masses and mixings,
is given in Appendix A of Ref.~\cite{Kang:2023iur}.
Here we repeat the details that are relevant for the present work.
With the addition of two $SU(2)_L$ scalar triplets, $\Delta_{1,2}$
and one inert $SU(2)_L$ doublet, $\Phi_2$,
to the SM Higgs $SU(2)_L$ doublet, $\Phi_1$,
the full scalar potential is given by
\begin{align}
	V = {}& V_\mathrm{IDM} + V_{\Delta}
		+ V_{H\Delta} + V_\mathrm{SB}\,, \\
	\label{eq:sclpotIDM}
	V_\mathrm{IDM} = {}& - m_{\Phi 1}^2 \Phi_1^\dagger \Phi_1 + m_{\Phi 2}^2 \Phi_2^\dagger \Phi_2 + \lambda_{\Phi 1} (\Phi_1^\dagger \Phi_1)^2
		+ \lambda_{\Phi 2} (\Phi_2^\dagger \Phi_2)^2 \nonumber \\
		& + \lambda_{\Phi 12} \Phi_1^\dagger \Phi_1 \Phi_2^\dagger \Phi_2 + \lambda'_{\Phi 12} \Phi_1^\dagger \Phi_2 \Phi_2^\dagger \Phi_1
		  + \lambda_5 \mathrm{Re}\left[(\Phi_1^\dagger \Phi_2)^2\right]\,, \\
	V_{\Delta} = {}& \sum_{n=1}^2 \left\{
            M_n^2 \Tr\left(\Delta_n^\dagger \Delta_n\right) + \lambda_{\Delta n} \Tr\left[\left(\Delta_n^\dagger \Delta_n\right)^2\right]
		 + \lambda'_{\Delta n} \left[\Tr\left(\Delta_n^\dagger \Delta_n\right)\right]^2\right\} \nonumber\\
		& + \lambda_{\Delta 12} \Tr(\Delta_1^\dagger \Delta_1 \Delta_2^\dagger \Delta_2)
		+ \lambda'_{\Delta 12} \Tr(\Delta_1^\dagger \Delta_1) \Tr(\Delta_2^\dagger \Delta_2) \nonumber\\
		& + \lambda_{\Delta 21} \Tr(\Delta_2^\dagger \Delta_1 \Delta_1^\dagger \Delta_2)
		+ \lambda'_{\Delta 21} \Tr(\Delta_2^\dagger \Delta_1) \Tr(\Delta_1^\dagger \Delta_2) \\
	V_{\Phi \Delta} = {}& \sum_{n=1}^{2}\sum_{k=1}^{2}\left[
            \lambda_{\Phi k\Delta n} \Phi_k^\dagger \Phi_k \Tr\left(\Delta_n^\dagger \Delta_n\right)
            + \lambda'_{\Phi k\Delta n} \Phi_k^\dagger \Delta_n  \Delta_n^\dagger \Phi_k
        \right]\,, \\
	V_\text{SB} = {}&
        \sum_{m=1}^2 \left\{
            \sum_{n=1}^2 \mu_{nm} \Phi_m^T i\sigma^2\Delta_n^\dagger \Phi_m
            +  \lambda_{\Phi k\Delta 12} \Phi_m^\dagger \Phi_m \Tr\left(\Delta_1^\dagger \Delta_2 \right)
		+ \lambda'_{\Phi k\Delta 12} \Phi_m^\dagger \Delta_1 \Delta_2^\dagger \Phi_m\right\} \nonumber \\
        & + \sum_{ijkl} \left[\lambda_{ijkl} \Tr\left(\Delta_i^\dagger \Delta_j \Delta_k^\dagger \Delta_l\right)
		+ \lambda'_{ijkl} \Tr\left(\Delta_i^\dagger \Delta_j\right)\left(\Delta_k^\dagger \Delta_l\right)\right]\nonumber\\
        & + M_{12}^2 \Tr\left(\Delta_1^\dagger \Delta_2\right) + \text{H.c.} \, ,
\end{align}
where we have also assumed an extra $Z_2$ symmetry,
under which all fields are assigned an even charge, except for $\Phi_2$.
The decomposition of the scalars is given in Eq.~(\ref{eq:scalars}).
The subscript SB in $V_\text{SB}$ stands for ``softly breaking''
since it softly breaks a global $U(1)$ symmetry.
In this part of the potential, the $(i,j,k,l)$ can take the values
(2,1,1,1), (1,2,1,1), (1,2,2,2), (2,1,2,2) and (1,2,1,2).
The VEV configuration of the scalars is given by
\begin{equation}
	\langle \Phi_1\rangle = \frac{1}{\sqrt{2}}\begin{pmatrix}
		0 \\
		v
	\end{pmatrix},\quad
	\langle\Delta_{n}\rangle = \frac{1}{\sqrt{2}}\begin{pmatrix}
		0 & 0 \\
		u_n & 0
	\end{pmatrix}\,,
\end{equation}
and we will reparameterize the $u_{1,2}$ VEVs
using $u_1^2 + u_2^2 = u^2$, $u_1 = u\cos\beta$ and $u_2 = u\sin\beta$~\cite{Ferreira:2021bdj}.
Considering that the VEVs must meet the condition $v^2 + 2u^2 \approx (246\ \text{GeV})^2$
and constraints on the $\rho$ parameter,
the value of $u$ cannot be above $\sim 8$~GeV.

In general,
it is possible to have mixings among neutral $CP$-even components, neutral $CP$-odd components,
the singly charged scalars, and the doubly charged scalars.
However, since $u \ll v$
and the couplings in $V_\text{SB}$ are assumed to be considerably smaller
than the couplings in the rest of the potential~\cite{tHooft:1979rat},
all the mixings become subleading.
Therefore, to simplify the expressions used in this work,
we have taken the \emph{no-mixing limit}
where the masses of the scalars are given by
\begin{align}
    m_h^2& = 2\lambda_{\Phi 1} v^2\,, \\
    \label{eq:mdelta1nomix}
    M_{\rho_1}^2 & =
        M_{\eta_1}^2 =
        M_{\delta^\pm_1}^2 =
        M_{\delta^{\pm\pm}_1}^2 =
        \frac{\mu_{11} v^2}{\sqrt{2} u \cos\beta} \equiv M_{\Delta_1}^2\,, \\
    \label{eq:mdelta2nomix}
    M_{\rho_2}^2 & =
        M_{\eta_2}^2 =
        M_{\delta^\pm_2}^2 =
        M_{\delta^{\pm\pm}_2}^2 =
        \frac{\mu_{21} v^2}{\sqrt{2} u \sin\beta} \equiv M_{\Delta_2}^2\,,
\end{align}
where $\rho_j$ and $\eta_j$ are the $CP$-even and $CP$-odd states in $\delta^0_j$.
For the dark scalars $H_0$, $A_0$ and $\Phi_2^{\pm}$,
their masses correspond to those of the original IDM.
A full development of the minimization of the potential,
masses and mixings
can be found in Ref.~\cite{Kang:2023iur}.

\bibliography{references}

\begin{thebibliography}{49}%
\makeatletter
\providecommand \@ifxundefined [1]{%
 \@ifx{#1\undefined}
}%
\providecommand \@ifnum [1]{%
 \ifnum #1\expandafter \@firstoftwo
 \else \expandafter \@secondoftwo
 \fi
}%
\providecommand \@ifx [1]{%
 \ifx #1\expandafter \@firstoftwo
 \else \expandafter \@secondoftwo
 \fi
}%
\providecommand \natexlab [1]{#1}%
\providecommand \enquote  [1]{``#1''}%
\providecommand \bibnamefont  [1]{#1}%
\providecommand \bibfnamefont [1]{#1}%
\providecommand \citenamefont [1]{#1}%
\providecommand \href@noop [0]{\@secondoftwo}%
\providecommand \href [0]{\begingroup \@sanitize@url \@href}%
\providecommand \@href[1]{\@@startlink{#1}\@@href}%
\providecommand \@@href[1]{\endgroup#1\@@endlink}%
\providecommand \@sanitize@url [0]{\catcode `\\12\catcode `\$12\catcode
  `\&12\catcode `\#12\catcode `\^12\catcode `\_12\catcode `\%12\relax}%
\providecommand \@@startlink[1]{}%
\providecommand \@@endlink[0]{}%
\providecommand \url  [0]{\begingroup\@sanitize@url \@url }%
\providecommand \@url [1]{\endgroup\@href {#1}{\urlprefix }}%
\providecommand \urlprefix  [0]{URL }%
\providecommand \Eprint [0]{\href }%
\providecommand \doibase [0]{https://doi.org/}%
\providecommand \selectlanguage [0]{\@gobble}%
\providecommand \bibinfo  [0]{\@secondoftwo}%
\providecommand \bibfield  [0]{\@secondoftwo}%
\providecommand \translation [1]{[#1]}%
\providecommand \BibitemOpen [0]{}%
\providecommand \bibitemStop [0]{}%
\providecommand \bibitemNoStop [0]{.\EOS\space}%
\providecommand \EOS [0]{\spacefactor3000\relax}%
\providecommand \BibitemShut  [1]{\csname bibitem#1\endcsname}%
\let\auto@bib@innerbib\@empty
\bibitem [{\citenamefont {Aghanim}\ \emph {et~al.}(2020)\citenamefont {Aghanim}
  \emph {et~al.}}]{Planck:2018vyg}%
  \BibitemOpen
  \bibfield  {author} {\bibinfo {author} {\bibfnamefont {N.}~\bibnamefont
  {Aghanim}} \emph {et~al.} (\bibinfo {collaboration} {Planck}),\ }\bibfield
  {title} {\bibinfo {title} {{Planck 2018 results. VI. Cosmological
  parameters}},\ }\href {https://doi.org/10.1051/0004-6361/201833910}
  {\bibfield  {journal} {\bibinfo  {journal} {Astron. Astrophys.}\ }\textbf
  {\bibinfo {volume} {641}},\ \bibinfo {pages} {A6} (\bibinfo {year} {2020})},\
  \bibinfo {note} {[Erratum: Astron.Astrophys. 652, C4 (2021)]},\ \Eprint
  {https://arxiv.org/abs/1807.06209} {arXiv:1807.06209 [astro-ph.CO]}
  \BibitemShut {NoStop}%
\bibitem [{\citenamefont {Sakharov}(1967)}]{Sakharov:1967dj}%
  \BibitemOpen
  \bibfield  {author} {\bibinfo {author} {\bibfnamefont {A.~D.}\ \bibnamefont
  {Sakharov}},\ }\bibfield  {title} {\bibinfo {title} {{Violation of CP
  Invariance, C asymmetry, and baryon asymmetry of the universe}},\ }\href
  {https://doi.org/10.1070/PU1991v034n05ABEH002497} {\bibfield  {journal}
  {\bibinfo  {journal} {Pisma Zh. Eksp. Teor. Fiz.}\ }\textbf {\bibinfo
  {volume} {5}},\ \bibinfo {pages} {32} (\bibinfo {year} {1967})}\BibitemShut
  {NoStop}%
\bibitem [{\citenamefont {Weinberg}(1979)}]{Weinberg:1979bt}%
  \BibitemOpen
  \bibfield  {author} {\bibinfo {author} {\bibfnamefont {S.}~\bibnamefont
  {Weinberg}},\ }\bibfield  {title} {\bibinfo {title} {{Cosmological Production
  of Baryons}},\ }\href {https://doi.org/10.1103/PhysRevLett.42.850} {\bibfield
   {journal} {\bibinfo  {journal} {Phys. Rev. Lett.}\ }\textbf {\bibinfo
  {volume} {42}},\ \bibinfo {pages} {850} (\bibinfo {year} {1979})}\BibitemShut
  {NoStop}%
\bibitem [{\citenamefont {Kolb}\ and\ \citenamefont
  {Wolfram}(1980)}]{Kolb:1979qa}%
  \BibitemOpen
  \bibfield  {author} {\bibinfo {author} {\bibfnamefont {E.~W.}\ \bibnamefont
  {Kolb}}\ and\ \bibinfo {author} {\bibfnamefont {S.}~\bibnamefont {Wolfram}},\
  }\bibfield  {title} {\bibinfo {title} {{Baryon Number Generation in the Early
  Universe}},\ }\href {https://doi.org/10.1016/0550-3213(82)90012-8} {\bibfield
   {journal} {\bibinfo  {journal} {Nucl. Phys. B}\ }\textbf {\bibinfo {volume}
  {172}},\ \bibinfo {pages} {224} (\bibinfo {year} {1980})},\ \bibinfo {note}
  {[Erratum: Nucl.Phys.B 195, 542 (1982)]}\BibitemShut {NoStop}%
\bibitem [{\citenamefont {Fukugita}\ and\ \citenamefont
  {Yanagida}(1986)}]{Fukugita:1986hr}%
  \BibitemOpen
  \bibfield  {author} {\bibinfo {author} {\bibfnamefont {M.}~\bibnamefont
  {Fukugita}}\ and\ \bibinfo {author} {\bibfnamefont {T.}~\bibnamefont
  {Yanagida}},\ }\bibfield  {title} {\bibinfo {title} {{Baryogenesis Without
  Grand Unification}},\ }\href {https://doi.org/10.1016/0370-2693(86)91126-3}
  {\bibfield  {journal} {\bibinfo  {journal} {Phys. Lett. B}\ }\textbf
  {\bibinfo {volume} {174}},\ \bibinfo {pages} {45} (\bibinfo {year}
  {1986})}\BibitemShut {NoStop}%
\bibitem [{\citenamefont {Workman}\ \emph {et~al.}(2022)\citenamefont {Workman}
  \emph {et~al.}}]{ParticleDataGroup:2022pth}%
  \BibitemOpen
  \bibfield  {author} {\bibinfo {author} {\bibfnamefont {R.~L.}\ \bibnamefont
  {Workman}} \emph {et~al.} (\bibinfo {collaboration} {Particle Data Group}),\
  }\bibfield  {title} {\bibinfo {title} {{Review of Particle Physics}},\ }\href
  {https://doi.org/10.1093/ptep/ptac097} {\bibfield  {journal} {\bibinfo
  {journal} {PTEP}\ }\textbf {\bibinfo {volume} {2022}},\ \bibinfo {pages}
  {083C01} (\bibinfo {year} {2022})}\BibitemShut {NoStop}%
\bibitem [{\citenamefont {Dasgupta}\ \emph {et~al.}(2020)\citenamefont
  {Dasgupta}, \citenamefont {Bhupal~Dev}, \citenamefont {Kang},\ and\
  \citenamefont {Zhang}}]{Dasgupta:2019lha}%
  \BibitemOpen
  \bibfield  {author} {\bibinfo {author} {\bibfnamefont {A.}~\bibnamefont
  {Dasgupta}}, \bibinfo {author} {\bibfnamefont {P.~S.}\ \bibnamefont
  {Bhupal~Dev}}, \bibinfo {author} {\bibfnamefont {S.~K.}\ \bibnamefont
  {Kang}},\ and\ \bibinfo {author} {\bibfnamefont {Y.}~\bibnamefont {Zhang}},\
  }\bibfield  {title} {\bibinfo {title} {{New mechanism for matter-antimatter
  asymmetry and connection with dark matter}},\ }\href
  {https://doi.org/10.1103/PhysRevD.102.055009} {\bibfield  {journal} {\bibinfo
   {journal} {Phys. Rev. D}\ }\textbf {\bibinfo {volume} {102}},\ \bibinfo
  {pages} {055009} (\bibinfo {year} {2020})},\ \Eprint
  {https://arxiv.org/abs/1911.03013} {arXiv:1911.03013 [hep-ph]} \BibitemShut
  {NoStop}%
\bibitem [{\citenamefont {Li}\ \emph {et~al.}(1985)\citenamefont {Li},
  \citenamefont {Liu},\ and\ \citenamefont {Wolfenstein}}]{Li:1985hy}%
  \BibitemOpen
  \bibfield  {author} {\bibinfo {author} {\bibfnamefont {L.-F.}\ \bibnamefont
  {Li}}, \bibinfo {author} {\bibfnamefont {Y.}~\bibnamefont {Liu}},\ and\
  \bibinfo {author} {\bibfnamefont {L.}~\bibnamefont {Wolfenstein}},\
  }\bibfield  {title} {\bibinfo {title} {Hidden {H}iggs particles},\ }\href
  {https://doi.org/10.1016/0370-2693(85)90117-0} {\bibfield  {journal}
  {\bibinfo  {journal} {Phys. Lett. B}\ }\textbf {\bibinfo {volume} {159}},\
  \bibinfo {pages} {45} (\bibinfo {year} {1985})}\BibitemShut {NoStop}%
\bibitem [{\citenamefont {Lusignoli}\ \emph {et~al.}(1990)\citenamefont
  {Lusignoli}, \citenamefont {Masiero},\ and\ \citenamefont
  {Roncadelli}}]{Lusignoli:1990yk}%
  \BibitemOpen
  \bibfield  {author} {\bibinfo {author} {\bibfnamefont {M.}~\bibnamefont
  {Lusignoli}}, \bibinfo {author} {\bibfnamefont {A.}~\bibnamefont {Masiero}},\
  and\ \bibinfo {author} {\bibfnamefont {M.}~\bibnamefont {Roncadelli}},\
  }\bibfield  {title} {\bibinfo {title} {{Spontaneous versus explicit breaking
  of a continuous global symmetry}},\ }\href
  {https://doi.org/10.1016/0370-2693(90)90868-7} {\bibfield  {journal}
  {\bibinfo  {journal} {Phys. Lett. B}\ }\textbf {\bibinfo {volume} {252}},\
  \bibinfo {pages} {247} (\bibinfo {year} {1990})}\BibitemShut {NoStop}%
\bibitem [{\citenamefont {de~S.~Pires}(2006)}]{deSPires:2005yok}%
  \BibitemOpen
  \bibfield  {author} {\bibinfo {author} {\bibfnamefont {C.~A.}\ \bibnamefont
  {de~S.~Pires}},\ }\bibfield  {title} {\bibinfo {title} {{Explicitly broken
  lepton number at low energy in the Higgs triplet model}},\ }\href
  {https://doi.org/10.1142/S0217732306019347} {\bibfield  {journal} {\bibinfo
  {journal} {Mod. Phys. Lett. A}\ }\textbf {\bibinfo {volume} {21}},\ \bibinfo
  {pages} {971} (\bibinfo {year} {2006})},\ \Eprint
  {https://arxiv.org/abs/hep-ph/0509152} {arXiv:hep-ph/0509152} \BibitemShut
  {NoStop}%
\bibitem [{\citenamefont {Freitas}\ \emph {et~al.}(2017)\citenamefont
  {Freitas}, \citenamefont {de~S.~Pires},\ and\ \citenamefont {Rodrigues~da
  Silva}}]{Freitas:2014fda}%
  \BibitemOpen
  \bibfield  {author} {\bibinfo {author} {\bibfnamefont {F.~F.}\ \bibnamefont
  {Freitas}}, \bibinfo {author} {\bibfnamefont {C.~A.}\ \bibnamefont
  {de~S.~Pires}},\ and\ \bibinfo {author} {\bibfnamefont {P.~S.}\ \bibnamefont
  {Rodrigues~da Silva}},\ }\bibfield  {title} {\bibinfo {title} {{Inverse type
  II seesaw mechanism and its signature at the LHC and ILC}},\ }\href
  {https://doi.org/10.1016/j.physletb.2017.03.016} {\bibfield  {journal}
  {\bibinfo  {journal} {Phys. Lett. B}\ }\textbf {\bibinfo {volume} {769}},\
  \bibinfo {pages} {48} (\bibinfo {year} {2017})},\ \Eprint
  {https://arxiv.org/abs/1408.5878} {arXiv:1408.5878 [hep-ph]} \BibitemShut
  {NoStop}%
\bibitem [{\citenamefont {de~Sousa~Pires}\ \emph {et~al.}(2019)\citenamefont
  {de~Sousa~Pires}, \citenamefont {Ferreira De~Freitas}, \citenamefont {Shu},
  \citenamefont {Huang},\ and\ \citenamefont {Wagner
  Vasconcelos~Oleg\'ario}}]{deSousaPires:2018fnl}%
  \BibitemOpen
  \bibfield  {author} {\bibinfo {author} {\bibfnamefont {C.~A.}\ \bibnamefont
  {de~Sousa~Pires}}, \bibinfo {author} {\bibfnamefont {F.}~\bibnamefont
  {Ferreira De~Freitas}}, \bibinfo {author} {\bibfnamefont {J.}~\bibnamefont
  {Shu}}, \bibinfo {author} {\bibfnamefont {L.}~\bibnamefont {Huang}},\ and\
  \bibinfo {author} {\bibfnamefont {P.}~\bibnamefont {Wagner
  Vasconcelos~Oleg\'ario}},\ }\bibfield  {title} {\bibinfo {title}
  {{Implementing the inverse type-II seesaw mechanism into the 3-3-1 model}},\
  }\href {https://doi.org/10.1016/j.physletb.2019.134827} {\bibfield  {journal}
  {\bibinfo  {journal} {Phys. Lett. B}\ }\textbf {\bibinfo {volume} {797}},\
  \bibinfo {pages} {134827} (\bibinfo {year} {2019})},\ \Eprint
  {https://arxiv.org/abs/1812.10570} {arXiv:1812.10570 [hep-ph]} \BibitemShut
  {NoStop}%
\bibitem [{\citenamefont {Deshpande}\ and\ \citenamefont
  {Ma}(1978)}]{Deshpande:1977rw}%
  \BibitemOpen
  \bibfield  {author} {\bibinfo {author} {\bibfnamefont {N.~G.}\ \bibnamefont
  {Deshpande}}\ and\ \bibinfo {author} {\bibfnamefont {E.}~\bibnamefont {Ma}},\
  }\bibfield  {title} {\bibinfo {title} {{Pattern of Symmetry Breaking with Two
  Higgs Doublets}},\ }\href {https://doi.org/10.1103/PhysRevD.18.2574}
  {\bibfield  {journal} {\bibinfo  {journal} {Phys. Rev. D}\ }\textbf {\bibinfo
  {volume} {18}},\ \bibinfo {pages} {2574} (\bibinfo {year}
  {1978})}\BibitemShut {NoStop}%
\bibitem [{\citenamefont {Lopez~Honorez}\ \emph {et~al.}(2007)\citenamefont
  {Lopez~Honorez}, \citenamefont {Nezri}, \citenamefont {Oliver},\ and\
  \citenamefont {Tytgat}}]{LopezHonorez:2006gr}%
  \BibitemOpen
  \bibfield  {author} {\bibinfo {author} {\bibfnamefont {L.}~\bibnamefont
  {Lopez~Honorez}}, \bibinfo {author} {\bibfnamefont {E.}~\bibnamefont
  {Nezri}}, \bibinfo {author} {\bibfnamefont {J.~F.}\ \bibnamefont {Oliver}},\
  and\ \bibinfo {author} {\bibfnamefont {M.~H.~G.}\ \bibnamefont {Tytgat}},\
  }\bibfield  {title} {\bibinfo {title} {{The Inert Doublet Model: An Archetype
  for Dark Matter}},\ }\href {https://doi.org/10.1088/1475-7516/2007/02/028}
  {\bibfield  {journal} {\bibinfo  {journal} {JCAP}\ }\textbf {\bibinfo
  {volume} {02}},\ \bibinfo {pages} {028}},\ \Eprint
  {https://arxiv.org/abs/hep-ph/0612275} {arXiv:hep-ph/0612275} \BibitemShut
  {NoStop}%
\bibitem [{\citenamefont {Gustafsson}(2010)}]{Gustafsson:2010zz}%
  \BibitemOpen
  \bibfield  {author} {\bibinfo {author} {\bibfnamefont {M.}~\bibnamefont
  {Gustafsson}},\ }\bibfield  {title} {\bibinfo {title} {{The Inert Doublet
  Model and Its Phenomenology}},\ }\href {https://doi.org/10.22323/1.114.0030}
  {\bibfield  {journal} {\bibinfo  {journal} {PoS}\ }\textbf {\bibinfo {volume}
  {CHARGED2010}},\ \bibinfo {pages} {030} (\bibinfo {year} {2010})},\ \Eprint
  {https://arxiv.org/abs/1106.1719} {arXiv:1106.1719 [hep-ph]} \BibitemShut
  {NoStop}%
\bibitem [{\citenamefont {Arhrib}\ \emph {et~al.}(2014)\citenamefont {Arhrib},
  \citenamefont {Tsai}, \citenamefont {Yuan},\ and\ \citenamefont
  {Yuan}}]{Arhrib:2013ela}%
  \BibitemOpen
  \bibfield  {author} {\bibinfo {author} {\bibfnamefont {A.}~\bibnamefont
  {Arhrib}}, \bibinfo {author} {\bibfnamefont {Y.-L.~S.}\ \bibnamefont {Tsai}},
  \bibinfo {author} {\bibfnamefont {Q.}~\bibnamefont {Yuan}},\ and\ \bibinfo
  {author} {\bibfnamefont {T.-C.}\ \bibnamefont {Yuan}},\ }\bibfield  {title}
  {\bibinfo {title} {{An Updated Analysis of Inert Higgs Doublet Model in light
  of the Recent Results from LUX, PLANCK, AMS-02 and LHC}},\ }\href
  {https://doi.org/10.1088/1475-7516/2014/06/030} {\bibfield  {journal}
  {\bibinfo  {journal} {JCAP}\ }\textbf {\bibinfo {volume} {06}},\ \bibinfo
  {pages} {030}},\ \Eprint {https://arxiv.org/abs/1310.0358} {arXiv:1310.0358
  [hep-ph]} \BibitemShut {NoStop}%
\bibitem [{\citenamefont {Belyaev}\ \emph {et~al.}(2018)\citenamefont
  {Belyaev}, \citenamefont {Cacciapaglia}, \citenamefont {Ivanov},
  \citenamefont {Rojas-Abatte},\ and\ \citenamefont
  {Thomas}}]{Belyaev:2016lok}%
  \BibitemOpen
  \bibfield  {author} {\bibinfo {author} {\bibfnamefont {A.}~\bibnamefont
  {Belyaev}}, \bibinfo {author} {\bibfnamefont {G.}~\bibnamefont
  {Cacciapaglia}}, \bibinfo {author} {\bibfnamefont {I.~P.}\ \bibnamefont
  {Ivanov}}, \bibinfo {author} {\bibfnamefont {F.}~\bibnamefont
  {Rojas-Abatte}},\ and\ \bibinfo {author} {\bibfnamefont {M.}~\bibnamefont
  {Thomas}},\ }\bibfield  {title} {\bibinfo {title} {{Anatomy of the Inert Two
  Higgs Doublet Model in the light of the LHC and non-LHC Dark Matter
  Searches}},\ }\href {https://doi.org/10.1103/PhysRevD.97.035011} {\bibfield
  {journal} {\bibinfo  {journal} {Phys. Rev. D}\ }\textbf {\bibinfo {volume}
  {97}},\ \bibinfo {pages} {035011} (\bibinfo {year} {2018})},\ \Eprint
  {https://arxiv.org/abs/1612.00511} {arXiv:1612.00511 [hep-ph]} \BibitemShut
  {NoStop}%
\bibitem [{\citenamefont {Fan}\ \emph {et~al.}(2022)\citenamefont {Fan},
  \citenamefont {Tang}, \citenamefont {Tsai},\ and\ \citenamefont
  {Wu}}]{Fan:2022dck}%
  \BibitemOpen
  \bibfield  {author} {\bibinfo {author} {\bibfnamefont {Y.-Z.}\ \bibnamefont
  {Fan}}, \bibinfo {author} {\bibfnamefont {T.-P.}\ \bibnamefont {Tang}},
  \bibinfo {author} {\bibfnamefont {Y.-L.~S.}\ \bibnamefont {Tsai}},\ and\
  \bibinfo {author} {\bibfnamefont {L.}~\bibnamefont {Wu}},\ }\bibfield
  {title} {\bibinfo {title} {{Inert Higgs Dark Matter for CDF II W-Boson Mass
  and Detection Prospects}},\ }\href
  {https://doi.org/10.1103/PhysRevLett.129.091802} {\bibfield  {journal}
  {\bibinfo  {journal} {Phys. Rev. Lett.}\ }\textbf {\bibinfo {volume} {129}},\
  \bibinfo {pages} {091802} (\bibinfo {year} {2022})},\ \Eprint
  {https://arxiv.org/abs/2204.03693} {arXiv:2204.03693 [hep-ph]} \BibitemShut
  {NoStop}%
\bibitem [{\citenamefont {Hambye}\ \emph {et~al.}(2001)\citenamefont {Hambye},
  \citenamefont {Ma},\ and\ \citenamefont {Sarkar}}]{Hambye:2000ui}%
  \BibitemOpen
  \bibfield  {author} {\bibinfo {author} {\bibfnamefont {T.}~\bibnamefont
  {Hambye}}, \bibinfo {author} {\bibfnamefont {E.}~\bibnamefont {Ma}},\ and\
  \bibinfo {author} {\bibfnamefont {U.}~\bibnamefont {Sarkar}},\ }\bibfield
  {title} {\bibinfo {title} {{Supersymmetric triplet Higgs model of neutrino
  masses and leptogenesis}},\ }\href
  {https://doi.org/10.1016/S0550-3213(01)00109-2} {\bibfield  {journal}
  {\bibinfo  {journal} {Nucl. Phys. B}\ }\textbf {\bibinfo {volume} {602}},\
  \bibinfo {pages} {23} (\bibinfo {year} {2001})},\ \Eprint
  {https://arxiv.org/abs/hep-ph/0011192} {arXiv:hep-ph/0011192} \BibitemShut
  {NoStop}%
\bibitem [{\citenamefont {Kang}\ and\ \citenamefont
  {Ramos}(2024)}]{Kang:2023iur}%
  \BibitemOpen
  \bibfield  {author} {\bibinfo {author} {\bibfnamefont {S.~K.}\ \bibnamefont
  {Kang}}\ and\ \bibinfo {author} {\bibfnamefont {R.}~\bibnamefont {Ramos}},\
  }\bibfield  {title} {\bibinfo {title} {{Common origin of dark matter, baryon
  asymmetry and neutrino masses in the standard model with extended scalars}},\
  }\href {https://doi.org/10.1007/s40042-024-01080-0} {\bibfield  {journal}
  {\bibinfo  {journal} {J. Korean Phys. Soc.}\ }\textbf {\bibinfo {volume}
  {85}},\ \bibinfo {pages} {1} (\bibinfo {year} {2024})},\ \Eprint
  {https://arxiv.org/abs/2309.08277} {arXiv:2309.08277 [hep-ph]} \BibitemShut
  {NoStop}%
\bibitem [{\citenamefont {Ferreira}\ \emph {et~al.}(2022)\citenamefont
  {Ferreira}, \citenamefont {Gon\c{c}alves},\ and\ \citenamefont
  {Joaquim}}]{Ferreira:2021bdj}%
  \BibitemOpen
  \bibfield  {author} {\bibinfo {author} {\bibfnamefont {P.~M.}\ \bibnamefont
  {Ferreira}}, \bibinfo {author} {\bibfnamefont {B.~L.}\ \bibnamefont
  {Gon\c{c}alves}},\ and\ \bibinfo {author} {\bibfnamefont {F.~R.}\
  \bibnamefont {Joaquim}},\ }\bibfield  {title} {\bibinfo {title} {{The hidden
  side of scalar-triplet models with spontaneous CP violation}},\ }\href
  {https://doi.org/10.1007/JHEP05(2022)105} {\bibfield  {journal} {\bibinfo
  {journal} {JHEP}\ }\textbf {\bibinfo {volume} {05}},\ \bibinfo {pages}
  {105}},\ \Eprint {https://arxiv.org/abs/2109.13179} {arXiv:2109.13179
  [hep-ph]} \BibitemShut {NoStop}%
\bibitem [{\citenamefont {Pich}\ and\ \citenamefont
  {Tuzon}(2009)}]{Pich:2009sp}%
  \BibitemOpen
  \bibfield  {author} {\bibinfo {author} {\bibfnamefont {A.}~\bibnamefont
  {Pich}}\ and\ \bibinfo {author} {\bibfnamefont {P.}~\bibnamefont {Tuzon}},\
  }\bibfield  {title} {\bibinfo {title} {{Yukawa Alignment in the
  Two-Higgs-Doublet Model}},\ }\href
  {https://doi.org/10.1103/PhysRevD.80.091702} {\bibfield  {journal} {\bibinfo
  {journal} {Phys. Rev. D}\ }\textbf {\bibinfo {volume} {80}},\ \bibinfo
  {pages} {091702} (\bibinfo {year} {2009})},\ \Eprint
  {https://arxiv.org/abs/0908.1554} {arXiv:0908.1554 [hep-ph]} \BibitemShut
  {NoStop}%
\bibitem [{\citenamefont {Navas}\ \emph {et~al.}(2024)\citenamefont {Navas}
  \emph {et~al.}}]{ParticleDataGroup:2024cfk}%
  \BibitemOpen
  \bibfield  {author} {\bibinfo {author} {\bibfnamefont {S.}~\bibnamefont
  {Navas}} \emph {et~al.} (\bibinfo {collaboration} {Particle Data Group}),\
  }\bibfield  {title} {\bibinfo {title} {{Review of particle physics}},\ }\href
  {https://doi.org/10.1103/PhysRevD.110.030001} {\bibfield  {journal} {\bibinfo
   {journal} {Phys. Rev. D}\ }\textbf {\bibinfo {volume} {110}},\ \bibinfo
  {pages} {030001} (\bibinfo {year} {2024})}\BibitemShut {NoStop}%
\bibitem [{\citenamefont {Kuzmin}\ \emph {et~al.}(1985)\citenamefont {Kuzmin},
  \citenamefont {Rubakov},\ and\ \citenamefont {Shaposhnikov}}]{Kuzmin:1985mm}%
  \BibitemOpen
  \bibfield  {author} {\bibinfo {author} {\bibfnamefont {V.~A.}\ \bibnamefont
  {Kuzmin}}, \bibinfo {author} {\bibfnamefont {V.~A.}\ \bibnamefont
  {Rubakov}},\ and\ \bibinfo {author} {\bibfnamefont {M.~E.}\ \bibnamefont
  {Shaposhnikov}},\ }\bibfield  {title} {\bibinfo {title} {{On the Anomalous
  Electroweak Baryon Number Nonconservation in the Early Universe}},\ }\href
  {https://doi.org/10.1016/0370-2693(85)91028-7} {\bibfield  {journal}
  {\bibinfo  {journal} {Phys. Lett. B}\ }\textbf {\bibinfo {volume} {155}},\
  \bibinfo {pages} {36} (\bibinfo {year} {1985})}\BibitemShut {NoStop}%
\bibitem [{\citenamefont {Harvey}\ and\ \citenamefont
  {Turner}(1990)}]{Harvey:1990qw}%
  \BibitemOpen
  \bibfield  {author} {\bibinfo {author} {\bibfnamefont {J.~A.}\ \bibnamefont
  {Harvey}}\ and\ \bibinfo {author} {\bibfnamefont {M.~S.}\ \bibnamefont
  {Turner}},\ }\bibfield  {title} {\bibinfo {title} {{Cosmological baryon and
  lepton number in the presence of electroweak fermion number violation}},\
  }\href {https://doi.org/10.1103/PhysRevD.42.3344} {\bibfield  {journal}
  {\bibinfo  {journal} {Phys. Rev. D}\ }\textbf {\bibinfo {volume} {42}},\
  \bibinfo {pages} {3344} (\bibinfo {year} {1990})}\BibitemShut {NoStop}%
\bibitem [{\citenamefont {D'Onofrio}\ \emph {et~al.}(2014)\citenamefont
  {D'Onofrio}, \citenamefont {Rummukainen},\ and\ \citenamefont
  {Tranberg}}]{DOnofrio:2014rug}%
  \BibitemOpen
  \bibfield  {author} {\bibinfo {author} {\bibfnamefont {M.}~\bibnamefont
  {D'Onofrio}}, \bibinfo {author} {\bibfnamefont {K.}~\bibnamefont
  {Rummukainen}},\ and\ \bibinfo {author} {\bibfnamefont {A.}~\bibnamefont
  {Tranberg}},\ }\bibfield  {title} {\bibinfo {title} {{Sphaleron Rate in the
  Minimal Standard Model}},\ }\href
  {https://doi.org/10.1103/PhysRevLett.113.141602} {\bibfield  {journal}
  {\bibinfo  {journal} {Phys. Rev. Lett.}\ }\textbf {\bibinfo {volume} {113}},\
  \bibinfo {pages} {141602} (\bibinfo {year} {2014})},\ \Eprint
  {https://arxiv.org/abs/1404.3565} {arXiv:1404.3565 [hep-ph]} \BibitemShut
  {NoStop}%
\bibitem [{\citenamefont {Pilaftsis}(1997)}]{Pilaftsis:1997dr}%
  \BibitemOpen
  \bibfield  {author} {\bibinfo {author} {\bibfnamefont {A.}~\bibnamefont
  {Pilaftsis}},\ }\bibfield  {title} {\bibinfo {title} {{Resonant CP violation
  induced by particle mixing in transition amplitudes}},\ }\href
  {https://doi.org/10.1016/S0550-3213(97)00469-0} {\bibfield  {journal}
  {\bibinfo  {journal} {Nucl. Phys. B}\ }\textbf {\bibinfo {volume} {504}},\
  \bibinfo {pages} {61} (\bibinfo {year} {1997})},\ \Eprint
  {https://arxiv.org/abs/hep-ph/9702393} {arXiv:hep-ph/9702393} \BibitemShut
  {NoStop}%
\bibitem [{\citenamefont {Roulet}\ \emph {et~al.}(1998)\citenamefont {Roulet},
  \citenamefont {Covi},\ and\ \citenamefont {Vissani}}]{Roulet:1997xa}%
  \BibitemOpen
  \bibfield  {author} {\bibinfo {author} {\bibfnamefont {E.}~\bibnamefont
  {Roulet}}, \bibinfo {author} {\bibfnamefont {L.}~\bibnamefont {Covi}},\ and\
  \bibinfo {author} {\bibfnamefont {F.}~\bibnamefont {Vissani}},\ }\bibfield
  {title} {\bibinfo {title} {{On the {CP} asymmetries in Majorana neutrino
  decays}},\ }\href {https://doi.org/10.1016/S0370-2693(98)00135-X} {\bibfield
  {journal} {\bibinfo  {journal} {Phys. Lett. B}\ }\textbf {\bibinfo {volume}
  {424}},\ \bibinfo {pages} {101} (\bibinfo {year} {1998})},\ \Eprint
  {https://arxiv.org/abs/hep-ph/9712468} {arXiv:hep-ph/9712468} \BibitemShut
  {NoStop}%
\bibitem [{\citenamefont {Pilaftsis}(1999)}]{Pilaftsis:1998pd}%
  \BibitemOpen
  \bibfield  {author} {\bibinfo {author} {\bibfnamefont {A.}~\bibnamefont
  {Pilaftsis}},\ }\bibfield  {title} {\bibinfo {title} {{Heavy Majorana
  neutrinos and baryogenesis}},\ }\href
  {https://doi.org/10.1142/S0217751X99000932} {\bibfield  {journal} {\bibinfo
  {journal} {Int. J. Mod. Phys. A}\ }\textbf {\bibinfo {volume} {14}},\
  \bibinfo {pages} {1811} (\bibinfo {year} {1999})},\ \Eprint
  {https://arxiv.org/abs/hep-ph/9812256} {arXiv:hep-ph/9812256} \BibitemShut
  {NoStop}%
\bibitem [{\citenamefont {Belyaev}\ \emph {et~al.}(2013)\citenamefont
  {Belyaev}, \citenamefont {Christensen},\ and\ \citenamefont
  {Pukhov}}]{Belyaev:2012qa}%
  \BibitemOpen
  \bibfield  {author} {\bibinfo {author} {\bibfnamefont {A.}~\bibnamefont
  {Belyaev}}, \bibinfo {author} {\bibfnamefont {N.~D.}\ \bibnamefont
  {Christensen}},\ and\ \bibinfo {author} {\bibfnamefont {A.}~\bibnamefont
  {Pukhov}},\ }\bibfield  {title} {\bibinfo {title} {{CalcHEP 3.4 for collider
  physics within and beyond the Standard Model}},\ }\href
  {https://doi.org/10.1016/j.cpc.2013.01.014} {\bibfield  {journal} {\bibinfo
  {journal} {Comput. Phys. Commun.}\ }\textbf {\bibinfo {volume} {184}},\
  \bibinfo {pages} {1729} (\bibinfo {year} {2013})},\ \Eprint
  {https://arxiv.org/abs/1207.6082} {arXiv:1207.6082 [hep-ph]} \BibitemShut
  {NoStop}%
\bibitem [{\citenamefont {Esteban}\ \emph {et~al.}(2024)\citenamefont
  {Esteban}, \citenamefont {Gonzalez-Garcia}, \citenamefont {Maltoni},
  \citenamefont {Martinez-Soler}, \citenamefont {Pinheiro},\ and\ \citenamefont
  {Schwetz}}]{Esteban:2024eli}%
  \BibitemOpen
  \bibfield  {author} {\bibinfo {author} {\bibfnamefont {I.}~\bibnamefont
  {Esteban}}, \bibinfo {author} {\bibfnamefont {M.~C.}\ \bibnamefont
  {Gonzalez-Garcia}}, \bibinfo {author} {\bibfnamefont {M.}~\bibnamefont
  {Maltoni}}, \bibinfo {author} {\bibfnamefont {I.}~\bibnamefont
  {Martinez-Soler}}, \bibinfo {author} {\bibfnamefont {J.~P.}\ \bibnamefont
  {Pinheiro}},\ and\ \bibinfo {author} {\bibfnamefont {T.}~\bibnamefont
  {Schwetz}},\ }\bibfield  {title} {\bibinfo {title} {{NuFit-6.0: updated
  global analysis of three-flavor neutrino oscillations}},\ }\href
  {https://doi.org/10.1007/JHEP12(2024)216} {\bibfield  {journal} {\bibinfo
  {journal} {JHEP}\ }\textbf {\bibinfo {volume} {12}},\ \bibinfo {pages}
  {216}},\ \Eprint {https://arxiv.org/abs/2410.05380} {arXiv:2410.05380
  [hep-ph]} \BibitemShut {NoStop}%
\bibitem [{\citenamefont {{NuFIT Collaboration}}(2024)}]{nufitwebsite}%
  \BibitemOpen
  \bibfield  {author} {\bibinfo {author} {\bibnamefont {{NuFIT
  Collaboration}}},\ }\href@noop {} {\bibinfo {title} {{NuFIT v6.0}}},\
  \bibinfo {howpublished} {\url{http://www.nu-fit.org/}} (\bibinfo {year}
  {2024})\BibitemShut {NoStop}%
\bibitem [{\citenamefont {{CMS Collaboration}}(2017)}]{CMS:2017pet}%
  \BibitemOpen
  \bibfield  {author} {\bibinfo {author} {\bibnamefont {{CMS Collaboration}}},\
  }\bibfield  {title} {\bibinfo {title} {{A search for doubly-charged Higgs
  boson production in three and four lepton final states at
  $\sqrt{s}=13~\mathrm{TeV}$}},\ }\href@noop {} {\bibfield  {journal} {\bibinfo
   {journal} {CMS-PAS-HIG-16-036}\ } (\bibinfo {year} {2017})}\BibitemShut
  {NoStop}%
\bibitem [{\citenamefont {Aaboud}\ \emph {et~al.}(2018)\citenamefont {Aaboud}
  \emph {et~al.}}]{ATLAS:2017xqs}%
  \BibitemOpen
  \bibfield  {author} {\bibinfo {author} {\bibfnamefont {M.}~\bibnamefont
  {Aaboud}} \emph {et~al.} (\bibinfo {collaboration} {ATLAS}),\ }\bibfield
  {title} {\bibinfo {title} {{Search for doubly charged Higgs boson production
  in multi-lepton final states with the ATLAS detector using
  proton\textendash{}proton collisions at $\sqrt{s}=13\,\text {TeV}$}},\ }\href
  {https://doi.org/10.1140/epjc/s10052-018-5661-z} {\bibfield  {journal}
  {\bibinfo  {journal} {Eur. Phys. J. C}\ }\textbf {\bibinfo {volume} {78}},\
  \bibinfo {pages} {199} (\bibinfo {year} {2018})},\ \Eprint
  {https://arxiv.org/abs/1710.09748} {arXiv:1710.09748 [hep-ex]} \BibitemShut
  {NoStop}%
\bibitem [{\citenamefont {Belanger}\ \emph {et~al.}(2021)\citenamefont
  {Belanger}, \citenamefont {Mjallal},\ and\ \citenamefont
  {Pukhov}}]{Belanger:2020gnr}%
  \BibitemOpen
  \bibfield  {author} {\bibinfo {author} {\bibfnamefont {G.}~\bibnamefont
  {Belanger}}, \bibinfo {author} {\bibfnamefont {A.}~\bibnamefont {Mjallal}},\
  and\ \bibinfo {author} {\bibfnamefont {A.}~\bibnamefont {Pukhov}},\
  }\bibfield  {title} {\bibinfo {title} {{Recasting direct detection limits
  within micrOMEGAs and implication for non-standard Dark Matter scenarios}},\
  }\href {https://doi.org/10.1140/epjc/s10052-021-09012-z} {\bibfield
  {journal} {\bibinfo  {journal} {Eur. Phys. J. C}\ }\textbf {\bibinfo {volume}
  {81}},\ \bibinfo {pages} {239} (\bibinfo {year} {2021})},\ \Eprint
  {https://arxiv.org/abs/2003.08621} {arXiv:2003.08621 [hep-ph]} \BibitemShut
  {NoStop}%
\bibitem [{\citenamefont {Aalbers}\ \emph {et~al.}(2025)\citenamefont {Aalbers}
  \emph {et~al.}}]{LZ:2024zvo}%
  \BibitemOpen
  \bibfield  {author} {\bibinfo {author} {\bibfnamefont {J.}~\bibnamefont
  {Aalbers}} \emph {et~al.} (\bibinfo {collaboration} {LZ}),\ }\bibfield
  {title} {\bibinfo {title} {{Dark Matter Search Results from
  4.2{\,}{\,}Tonne-Years of Exposure of the LUX-ZEPLIN (LZ) Experiment}},\
  }\href {https://doi.org/10.1103/4dyc-z8zf} {\bibfield  {journal} {\bibinfo
  {journal} {Phys. Rev. Lett.}\ }\textbf {\bibinfo {volume} {135}},\ \bibinfo
  {pages} {011802} (\bibinfo {year} {2025})},\ \Eprint
  {https://arxiv.org/abs/2410.17036} {arXiv:2410.17036 [hep-ex]} \BibitemShut
  {NoStop}%
\bibitem [{\citenamefont {Abdalla}\ \emph {et~al.}(2022)\citenamefont {Abdalla}
  \emph {et~al.}}]{HESS:2022ygk}%
  \BibitemOpen
  \bibfield  {author} {\bibinfo {author} {\bibfnamefont {H.}~\bibnamefont
  {Abdalla}} \emph {et~al.} (\bibinfo {collaboration} {H.E.S.S.}),\ }\bibfield
  {title} {\bibinfo {title} {{Search for Dark Matter Annihilation Signals in
  the H.E.S.S. Inner Galaxy Survey}},\ }\href
  {https://doi.org/10.1103/PhysRevLett.129.111101} {\bibfield  {journal}
  {\bibinfo  {journal} {Phys. Rev. Lett.}\ }\textbf {\bibinfo {volume} {129}},\
  \bibinfo {pages} {111101} (\bibinfo {year} {2022})},\ \Eprint
  {https://arxiv.org/abs/2207.10471} {arXiv:2207.10471 [astro-ph.HE]}
  \BibitemShut {NoStop}%
\bibitem [{\citenamefont {Lebedev}\ and\ \citenamefont
  {Lee}(2011)}]{Lebedev:2011aq}%
  \BibitemOpen
  \bibfield  {author} {\bibinfo {author} {\bibfnamefont {O.}~\bibnamefont
  {Lebedev}}\ and\ \bibinfo {author} {\bibfnamefont {H.~M.}\ \bibnamefont
  {Lee}},\ }\bibfield  {title} {\bibinfo {title} {{Higgs Portal Inflation}},\
  }\href {https://doi.org/10.1140/epjc/s10052-011-1821-0} {\bibfield  {journal}
  {\bibinfo  {journal} {Eur. Phys. J. C}\ }\textbf {\bibinfo {volume} {71}},\
  \bibinfo {pages} {1821} (\bibinfo {year} {2011})},\ \Eprint
  {https://arxiv.org/abs/1105.2284} {arXiv:1105.2284 [hep-ph]} \BibitemShut
  {NoStop}%
\bibitem [{\citenamefont {Starobinsky}(1979)}]{Starobinsky:1979ty}%
  \BibitemOpen
  \bibfield  {author} {\bibinfo {author} {\bibfnamefont {A.~A.}\ \bibnamefont
  {Starobinsky}},\ }\bibfield  {title} {\bibinfo {title} {{Spectrum of relict
  gravitational radiation and the early state of the universe}},\ }\href@noop
  {} {\bibfield  {journal} {\bibinfo  {journal} {JETP Lett.}\ }\textbf
  {\bibinfo {volume} {30}},\ \bibinfo {pages} {682} (\bibinfo {year}
  {1979})}\BibitemShut {NoStop}%
\bibitem [{\citenamefont {Giudice}\ and\ \citenamefont
  {Lee}(2011)}]{Giudice:2010ka}%
  \BibitemOpen
  \bibfield  {author} {\bibinfo {author} {\bibfnamefont {G.~F.}\ \bibnamefont
  {Giudice}}\ and\ \bibinfo {author} {\bibfnamefont {H.~M.}\ \bibnamefont
  {Lee}},\ }\bibfield  {title} {\bibinfo {title} {{Unitarizing Higgs
  Inflation}},\ }\href {https://doi.org/10.1016/j.physletb.2010.10.035}
  {\bibfield  {journal} {\bibinfo  {journal} {Phys. Lett. B}\ }\textbf
  {\bibinfo {volume} {694}},\ \bibinfo {pages} {294} (\bibinfo {year}
  {2011})},\ \Eprint {https://arxiv.org/abs/1010.1417} {arXiv:1010.1417
  [hep-ph]} \BibitemShut {NoStop}%
\bibitem [{\citenamefont {Das}\ and\ \citenamefont {Khan}(2023)}]{Das:2022qyc}%
  \BibitemOpen
  \bibfield  {author} {\bibinfo {author} {\bibfnamefont {P.}~\bibnamefont
  {Das}}\ and\ \bibinfo {author} {\bibfnamefont {N.}~\bibnamefont {Khan}},\
  }\bibfield  {title} {\bibinfo {title} {{Origin of neutrino masses, dark
  matter, leptogenesis, and inflation in a seesaw model with triplets}},\
  }\href {https://doi.org/10.1103/PhysRevD.107.075008} {\bibfield  {journal}
  {\bibinfo  {journal} {Phys. Rev. D}\ }\textbf {\bibinfo {volume} {107}},\
  \bibinfo {pages} {075008} (\bibinfo {year} {2023})},\ \Eprint
  {https://arxiv.org/abs/2207.01238} {arXiv:2207.01238 [hep-ph]} \BibitemShut
  {NoStop}%
\bibitem [{\citenamefont {Bezrukov}\ \emph {et~al.}(2018)\citenamefont
  {Bezrukov}, \citenamefont {Pauly},\ and\ \citenamefont
  {Rubio}}]{Bezrukov:2017dyv}%
  \BibitemOpen
  \bibfield  {author} {\bibinfo {author} {\bibfnamefont {F.}~\bibnamefont
  {Bezrukov}}, \bibinfo {author} {\bibfnamefont {M.}~\bibnamefont {Pauly}},\
  and\ \bibinfo {author} {\bibfnamefont {J.}~\bibnamefont {Rubio}},\ }\bibfield
   {title} {\bibinfo {title} {{On the robustness of the primordial power
  spectrum in renormalized Higgs inflation}},\ }\href
  {https://doi.org/10.1088/1475-7516/2018/02/040} {\bibfield  {journal}
  {\bibinfo  {journal} {JCAP}\ }\textbf {\bibinfo {volume} {02}},\ \bibinfo
  {pages} {040}},\ \Eprint {https://arxiv.org/abs/1706.05007} {arXiv:1706.05007
  [hep-ph]} \BibitemShut {NoStop}%
\bibitem [{\citenamefont {Allahverdi}\ \emph {et~al.}(2010)\citenamefont
  {Allahverdi}, \citenamefont {Brandenberger}, \citenamefont {Cyr-Racine},\
  and\ \citenamefont {Mazumdar}}]{Allahverdi:2010xz}%
  \BibitemOpen
  \bibfield  {author} {\bibinfo {author} {\bibfnamefont {R.}~\bibnamefont
  {Allahverdi}}, \bibinfo {author} {\bibfnamefont {R.}~\bibnamefont
  {Brandenberger}}, \bibinfo {author} {\bibfnamefont {F.-Y.}\ \bibnamefont
  {Cyr-Racine}},\ and\ \bibinfo {author} {\bibfnamefont {A.}~\bibnamefont
  {Mazumdar}},\ }\bibfield  {title} {\bibinfo {title} {{Reheating in
  Inflationary Cosmology: Theory and Applications}},\ }\href
  {https://doi.org/10.1146/annurev.nucl.012809.104511} {\bibfield  {journal}
  {\bibinfo  {journal} {Ann. Rev. Nucl. Part. Sci.}\ }\textbf {\bibinfo
  {volume} {60}},\ \bibinfo {pages} {27} (\bibinfo {year} {2010})},\ \Eprint
  {https://arxiv.org/abs/1001.2600} {arXiv:1001.2600 [hep-th]} \BibitemShut
  {NoStop}%
\bibitem [{\citenamefont {Bezrukov}\ \emph {et~al.}(2009)\citenamefont
  {Bezrukov}, \citenamefont {Gorbunov},\ and\ \citenamefont
  {Shaposhnikov}}]{Bezrukov:2008ut}%
  \BibitemOpen
  \bibfield  {author} {\bibinfo {author} {\bibfnamefont {F.}~\bibnamefont
  {Bezrukov}}, \bibinfo {author} {\bibfnamefont {D.}~\bibnamefont {Gorbunov}},\
  and\ \bibinfo {author} {\bibfnamefont {M.}~\bibnamefont {Shaposhnikov}},\
  }\bibfield  {title} {\bibinfo {title} {{On initial conditions for the Hot Big
  Bang}},\ }\href {https://doi.org/10.1088/1475-7516/2009/06/029} {\bibfield
  {journal} {\bibinfo  {journal} {JCAP}\ }\textbf {\bibinfo {volume} {06}},\
  \bibinfo {pages} {029}},\ \Eprint {https://arxiv.org/abs/0812.3622}
  {arXiv:0812.3622 [hep-ph]} \BibitemShut {NoStop}%
\bibitem [{\citenamefont {Linde}(1982)}]{Linde:1981mu}%
  \BibitemOpen
  \bibfield  {author} {\bibinfo {author} {\bibfnamefont {A.~D.}\ \bibnamefont
  {Linde}},\ }\bibfield  {title} {\bibinfo {title} {{A New Inflationary
  Universe Scenario: A Possible Solution of the Horizon, Flatness, Homogeneity,
  Isotropy and Primordial Monopole Problems}},\ }\href
  {https://doi.org/10.1016/0370-2693(82)91219-9} {\bibfield  {journal}
  {\bibinfo  {journal} {Phys. Lett. B}\ }\textbf {\bibinfo {volume} {108}},\
  \bibinfo {pages} {389} (\bibinfo {year} {1982})}\BibitemShut {NoStop}%
\bibitem [{\citenamefont {Choubey}\ and\ \citenamefont
  {Kumar}(2017)}]{Choubey:2017hsq}%
  \BibitemOpen
  \bibfield  {author} {\bibinfo {author} {\bibfnamefont {S.}~\bibnamefont
  {Choubey}}\ and\ \bibinfo {author} {\bibfnamefont {A.}~\bibnamefont
  {Kumar}},\ }\bibfield  {title} {\bibinfo {title} {{Inflation and Dark Matter
  in the Inert Doublet Model}},\ }\href
  {https://doi.org/10.1007/JHEP11(2017)080} {\bibfield  {journal} {\bibinfo
  {journal} {JHEP}\ }\textbf {\bibinfo {volume} {11}},\ \bibinfo {pages}
  {080}},\ \Eprint {https://arxiv.org/abs/1707.06587} {arXiv:1707.06587
  [hep-ph]} \BibitemShut {NoStop}%
\bibitem [{\citenamefont {Billard}\ \emph {et~al.}(2022)\citenamefont {Billard}
  \emph {et~al.}}]{Billard:2021uyg}%
  \BibitemOpen
  \bibfield  {author} {\bibinfo {author} {\bibfnamefont {J.}~\bibnamefont
  {Billard}} \emph {et~al.},\ }\bibfield  {title} {\bibinfo {title} {{Direct
  detection of dark matter\textemdash{}APPEC committee report*}},\ }\href
  {https://doi.org/10.1088/1361-6633/ac5754} {\bibfield  {journal} {\bibinfo
  {journal} {Rept. Prog. Phys.}\ }\textbf {\bibinfo {volume} {85}},\ \bibinfo
  {pages} {056201} (\bibinfo {year} {2022})},\ \Eprint
  {https://arxiv.org/abs/2104.07634} {arXiv:2104.07634 [hep-ex]} \BibitemShut
  {NoStop}%
\bibitem [{\citenamefont {de~Melo}\ \emph {et~al.}(2019)\citenamefont
  {de~Melo}, \citenamefont {Queiroz},\ and\ \citenamefont
  {Villamizar}}]{deMelo:2019asm}%
  \BibitemOpen
  \bibfield  {author} {\bibinfo {author} {\bibfnamefont {T.~B.}\ \bibnamefont
  {de~Melo}}, \bibinfo {author} {\bibfnamefont {F.~S.}\ \bibnamefont
  {Queiroz}},\ and\ \bibinfo {author} {\bibfnamefont {Y.}~\bibnamefont
  {Villamizar}},\ }\bibfield  {title} {\bibinfo {title} {{Doubly Charged Scalar
  at the High-Luminosity and High-Energy LHC}},\ }\href
  {https://doi.org/10.1142/S0217751X19501574} {\bibfield  {journal} {\bibinfo
  {journal} {Int. J. Mod. Phys. A}\ }\textbf {\bibinfo {volume} {34}},\
  \bibinfo {pages} {1950157} (\bibinfo {year} {2019})},\ \Eprint
  {https://arxiv.org/abs/1909.07429} {arXiv:1909.07429 [hep-ph]} \BibitemShut
  {NoStop}%
\bibitem [{\citenamefont {'t~Hooft}(1980)}]{tHooft:1979rat}%
  \BibitemOpen
  \bibfield  {author} {\bibinfo {author} {\bibfnamefont {G.}~\bibnamefont
  {'t~Hooft}},\ }\bibfield  {title} {\bibinfo {title} {{Naturalness, chiral
  symmetry, and spontaneous chiral symmetry breaking}},\ }\href
  {https://doi.org/10.1007/978-1-4684-7571-5_9} {\bibfield  {journal} {\bibinfo
   {journal} {NATO Sci. Ser. B}\ }\textbf {\bibinfo {volume} {59}},\ \bibinfo
  {pages} {135} (\bibinfo {year} {1980})}\BibitemShut {NoStop}%
\end{thebibliography}%

\end{document}